\documentclass[journal]{IEEEtran}

\usepackage{times}
\usepackage{epsfig}
\usepackage{graphicx}
\usepackage{amsmath}
\usepackage{amssymb}
\usepackage[ruled,vlined]{algorithm2e}
\usepackage{algpseudocode}
\usepackage{caption}
\usepackage{subcaption}
\usepackage{makecell}
\usepackage{booktabs}
\usepackage{tabularx}
\usepackage{xcolor}
\usepackage{bm}
\usepackage{cite}
\usepackage{textcomp}
\usepackage{mathrsfs}
\usepackage{graphicx}
\graphicspath{ {./images/} }

\begin{document}
\title{Coordinated FMCW and OFDM for Integrated Sensing and Communication}

\author{Yuhong Wang, Yonghong Zeng,~\IEEEmembership{Fellow,~IEEE,} Sumei Sun,~\IEEEmembership{Fellow,~IEEE}, and Xiaojuan Zhang,~\IEEEmembership{Member,~IEEE}
\thanks{This research is supported by the National Research Foundation, Singapore, and Infocomm Media Development Authority under its Future Communications Research and Development Program. (Corresponding author: Yonghong Zeng.)}
\thanks{The authors are with the Institute for Infocomm Research (${\rm I}^2$R), Agency for Science, Technology and Research (${\rm A}^*$STAR), Singapore. Email: \{wangyh, yhzeng, sunsm, xzhang\}@i2r.a-star.edu.sg.}}

\maketitle

\begin{abstract}
Modern vehicles commonly use Frequency Modulated Continuous Wave (FMCW) radar for environmental sensing and Orthogonal Frequency Division Multiplexing (OFDM) for wireless communication. However, operating these systems independently leads to increased hardware complexity, cost, and inefficient spectrum usage. To address this, we propose a coordinated FMCW-OFDM (Co-FMCW-OFDM) system that enables integrated sensing and communication (ISAC) by allowing sensing and communication to share the same RF front end, antennas, and spectral resources.
In the proposed ISAC system, the FMCW signal is superimposed on the OFDM signal and serves dual purposes: facilitating bistatic sensing and enabling channel estimation at the receiver end. Based on proposed Co-FMCW-OFDM waveform, we propose two efficient sensing algorithms—fast cyclic correlation radar (FCCR) and digital mixing and down-sampling (DMD)—which significantly reduce system complexity while accurately estimating target range and velocity. 
We consider a realistic channel model where delays can take any value, not just integer multiples of the sampling period. This leads to a significantly larger number of effective paths compared to the actual number of targets, which makes the sensing, channel estimation, and interference cancellation more challenging. 
Leveraging the sensing results, we develop a sensing-aided effective channel estimation method which effectively reconstructs the channel under arbitrary delay condition based on successive interference cancellation and propose an interference cancellation scheme that removes the FMCW signal before the OFDM demodulation. Simulation results demonstrate that the proposed system achieves superior sensing accuracy, improved channel estimation, and lower bit error rate (BER) compared to conventional OFDM systems with embedded pilots. The proposed scheme demonstrates superior BER performance in comparison to the conventional OFDM-plus-FMCW approach.

\end{abstract}



\section{Introduction} \label{introduction}

Modern vehicles commonly use Frequency Modulated Continuous Wave (FMCW) radar for environmental sensing and Orthogonal Frequency Division Multiplexing (OFDM) for wireless communication. However, operating these systems independently leads to increased hardware complexity, cost, and inefficient spectrum usage. By sharing spectrum, hardware and software, integrated sensing and communication (ISAC) can enhance spectrum efficiency, reduce
device size, lower power consumption, and decrease costs for both
communication and sensing functions. As a result, ISAC has been
envisioned as a key enabler for a number of emerging applications, such
as intelligent transportation systems, smart drones, health care and
home security \cite{LiuFan2020-1,LiuFan2022-2,Hexa-X2023-13}.

\subsection{Background and motivation}

One of the main challenges in ISAC lies in finding suitable waveforms that can be simultaneously used for information transmission and radar sensing. In recent years, there have been a lot of researches for ISAC waveform design. The design of ISAC waveforms encompasses a wide range of strategies that balance or prioritize sensing and communication objectives depending on the application context. In general, these strategies can be categorized into three paradigms: (1) Communication-Centric Design: Prioritizes communication performance while embedding sensing functionality with minimal impact. Typical approaches reuse existing communication signals (e.g. OFDM) for sensing with slight modifications \cite{XTian2017-OFDM1}. (2) Sensing-Centric Design: Focuses on maintaining high-precision sensing capabilities while enabling basic communication features. These designs often originate from radar systems and embed data within radar waveforms \cite{MJNowak2016-fmcw1}. (3) Joint or Balanced Design: Aims to co-optimize both sensing and communication performances, often formulated as multi-objective optimization problems. This includes, but is not limited to, weighted-sum utility functions that combine metrics such as communication rate and sensing accuracy. Optimization can be done in the time-frequency domain through power allocation \cite{Keskin2021-19,LiuY2017-20,Keskin2020-21} or in the spatial domain through precoding \cite{LiChen2021-22,Haocheng2023,BinLiao2023,XiangLiu2020,XinHe2020,Yingdu2023,LiuFan2021-17,LiuFan2018-24}. One of the major disadvantages of such approaches is that the designd waveform is only optimized for the specified environment/channel. Furthermore, such designs in general have very high complexity in solving the optimization problems. 

Orthogonal time frequency space (OTFS) and its extensions are promising candidates for ISAC \cite{otfs-2017,PRav2019,Linhai-2022,Farhang-2023-ICC,Tusha-2023,Zhang-2023-ICC,Dazhuo2024}. Research has shown that OTFS can be used as a dual purpose waveform to enhance communication performance in high mobility environment \cite{otfs-2017,PRav2019,Linhai-2022,Farhang-2023-ICC,Tusha-2023} and enable reliable sensing for ISAC \cite{Zhang-2023-ICC,Dazhuo2024}. However, one of the major obstacles for OTFS is the high complexity receiver. 

Orthogonal frequency division multiplexing (OFDM) is the waveform standardized for the fifth generation (5G) and the fourth generation (4G) communications, which enjoys a low complexity receiver to combat the channel frequency selectivity and enables a flexible multiple access scheme (OFDMA). Adopting OFDM as the ISAC waveform for both communication and sensing purposes
has been extensively studied in recent years \cite{Sturm2009-3,Baquero2019-4,Nata2023-5,Zhang2023-6}.
Ref.~\cite{Sturm2009-3} introduces a channel estimation-based sensing
algorithm using OFDM signals. Ref.~\cite{Baquero2019-4} explores the use of
5G reference signals (RS) for bistatic sensing. Ref.~\cite{Nata2023-5} investigates a bistatic radar sensing system that utilizes 5G signals by combining different reference signals and interpolating between nonuniform grid points to reconstruct
the channel for the entire grid. It shows that leveraging decision-feedback from decoded payload data as ``pseudo-RS'' can improve delay-Doppler detectability and robustness to noise. Additionally, ref.~\cite{Zhang2023-6} implements a bistatic radar based on standardized 5G waveforms, using decision feedback from decoded payload data as ``pseudo-RS'' combined with demodulation reference signals (DMRS) to estimate the range, speed, and angle of radar targets.

FMCW radar and OFDM communication devices
are widely used in various applications. For instance, in advanced
automobiles, FMCW radar is typically employed for environmental sensing,
while OFDM communication devices are used to connect with other vehicles
or infrastructure. Integrated designs for both systems have the
potential to reduce complexity and cost, share spectrum, and enhance
performance. Thus, we propose a coordinated FMCW and OFDM (Co-FMCW-OFDM), in which FMCW signal and OFDM signal are designed to be coordinated and synchronized and share the same RF front end, antenna and the spectral resources. 

Our motivation is to explore a practical and low-complexity ISAC solution that can be rapidly deployed using existing vehicular hardware without major architectural changes. This addresses a critical need in real-world automotive applications where cost, legacy compatibility, and spectrum efficiency are key constraints. Our approach maximizes the utilization of existing vehicular hardware, thereby facilitating the rapid and practical deployment of ISAC systems.

In contrast to most existing ISAC solutions—which are typically based on either communication-centric waveforms, sensing-centric waveforms, or jointly optimized designs—our approach emphasizes practicality and low complexity. Existing solutions often suffer from trade-offs such as degraded radar accuracy, reduced communication robustness, excessive implementation complexity, limited adaptability to varying environments, or reliance on custom RF front-end architectures.

In the proposed framework, FMCW signals serve dual purposes: facilitating sensing and enabling channel estimation at the receiver end. Different from most papers, which only consider integer delays (path delays are integer multiples of sampling period), we consider practical situations where the delay can have any value. Based on the characteristics of the FMCW signal, we propose two methods to estimate the delays and Doppler frequencies for all paths: (1)  digital
mixing and down-sampling (DMD), (2) fast cyclic correlation radar
(FCCR) \cite{Zeng2020-7}. With the estimated delays and Doppler frequencies, we propose a sensing-aided channel estimation algorithm with successive interference cancellation (SIC), which achieves super performance. Since the FMCW signal is superimposed with the OFDM signal, it is necessary to remove the FMCW
signal interference before demodulating the OFDM signal to ensure
optimal demodulation performance. By using the information obtained from
sensing and channel estimation, the interference caused by FMCW signals
is reconstructed and eliminated from the received signal prior to OFDM
demodulation. 

There have been substantial researches on the combination of FMCW and OFDM signals for ISAC, as seen in \cite{Wang2024-8, Arslan2020, ArslanOpenJournal2020, Arslan2022, Aydogdu2019TransIT, Aydogdu2020SigProcMgz}. For example, Ref. \cite{Wang2024-8} proposed a power-domain non-orthogonal ISAC waveform using FMCW and OFDM, and analyzed the impact of sensing signal discretization on overall system performance. Our proposed scheme differs from  \cite{Wang2024-8} in several key aspects:
(1)	Ref \cite{Wang2024-8}   employs a non-zero cyclic prefix (CP) for the FMCW signal. In contrast, we adopt an all-zero CP for the FMCW signal, which is inherited from the FMCW radar waveform design with guard interval, and reduces interference to the overlapping OFDM symbols. 
(2)	Ref \cite{Wang2024-8} does not leverage the FMCW signal for channel estimation and does not propose any method for estimating the channel. Our scheme utilizes the FMCW signal for both sensing and channel estimation. Specifically, we propose a sensing-aided channel estimation method based on SIC, which effectively improves performance in dynamic channel conditions.
(3)	Ref \cite{Wang2024-8} performs interference cancellation after OFDM demodulation and equalization in the frequency domain, which uses the averaged received signal as the estimation of the FMCW signal. It then subtracts the estimated FMCW signal from the received signal. In our approach, interference cancellation is performed in the time domain, prior to OFDM demodulation. We use the estimated channel and the known FMCW signal to cancel the FMCW contribution, which is more accurate and yields better performance. Simulation results demonstrate that this method achieves superior interference suppression and significantly enhances overall system performance.

Ref \cite{Arslan2020} and \cite{ArslanOpenJournal2020} proposed a non-orthogonal superposition of FMCW and OFDM signals. In their scheme, the transmitter is coordinated such that the first chirp of each transmission frame is reserved solely for sensing—no OFDM data is transmitted during this chirp. While this avoids mutual interference, it also reduces communication data rate. In contrast, our proposed scheme transmits FMCW and OFDM signals simultaneously, maintaining full data throughput.
Ref \cite{Arslan2022} introduced a waveform design where FMCW and OFDM coexist orthogonally on the same time-frequency resources. This orthogonality is achieved by constraining the FMCW chirp duration to align with pilot subcarriers in the frequency domain, enabling channel estimation. However, this constraint limits the flexibility of the FMCW waveform, potentially compromising radar sensing performance.
Ref \cite{Aydogdu2019TransIT} and \cite{Aydogdu2020SigProcMgz} explored various interference mitigation strategies, such as time and frequency resource coordination, to enable the coexistence of FMCW and OFDM waveforms.

\subsection{Our Contributions}
The following are the main contributions of our work.

\begin{itemize}
\item
  We propose a coordinated FMCW and OFDM waveform for ISAC. In this scheme, the FMCW radar signal is superimposed on the OFDM communication signal and acts as a pilot for both bistatic sensing and channel estimation at the receiver. FMCW signal and OFDM signal have the same symbol duration and are synchronized in symbol level. A special feature of our design is the adoption of an all-zero cyclic prefix (CP) for the FMCW signal. This has two primary benefits: Firstly, it enables the use of efficient and low-complexity algorithms FCCR and DMD for sensing; Secondly, it reduces the interference from the FMCW signal to the overlapping OFDM symbols, thereby enhancing communication performance.
\item
We propose two bistatic sensing algorithms, FCCR and DMD, for estimating the range and velocity of targets using FMCW signals. In contrast to traditional FMCW radar systems, where the received signal undergoes analogue matched filtering (mixing with the transmitted analogue FMCW signal), low-pass filtering, and sampling, the proposed DMD algorithm performs digital mixing and down-sampling based on the digitized FMCW signal.  The FCCR algorithm, on the other hand, has been shown to be near-optimal for radar sensing of signals with block structures, such as cyclic-prefixed single-carrier (CP-SC) and CP-OFDM waveforms. Thanks to the specific structure of our FMCW signal design, the FCCR algorithm can be directly applied. 

\item Unlike many prior works that assume delays are integer multiples of the sampling period, we consider a realistic channel model where delays can take  arbitrary value, not just integer multiples of the sampling period. This leads to a significantly larger number of effective paths compared to the actual number of targets, which makes the sensing and channel estimation more challenging. We derive a signal model which captures the spreading of energy across multiple taps, increasing the complexity of sensing and estimation. This modelling is more aligned with real-world propagation environments, especially for high-resolution sensing applications, yet has received limited attention in existing ISAC literature. 
Our contribution lies in both deriving a mathematically tractable signal model under arbitrary delays and highlighting the implications for sensing-channel estimation interaction and interference cancellation.

\item Leveraging delay and Doppler frequency estimation from the sensing module, we propose a sensing-aided effective channel estimation method which effectively reconstructs the channel under arbitrary delay condition based on SIC. This method accommodates the arbitrary delays of the effective channel and iteratively refines channel estimates. While SIC is a known technique, its tailored application for sensing-informed estimation in the presence of arbitrary delays—within the Co-FMCW-OFDM framework—is a novel adaptation. We also provide a performance analysis of the channel estimator, which is typically lacking in related works.

\item We develop and analyze interference cancellation methods to mitigate the impact of FMCW components on OFDM demodulation. We present a method that reconstructs the FMCW signal component using the joint sensing-channel estimates and then removes it in the time domain prior to OFDM demodulation.
Unlike conventional interference cancellation methods which cancel FMCW interference in frequency domain \cite{Wang2024-8} , our method explicitly models the FMCW signal structure and integrates it with real-time sensing results to achieve time domain FMCW signal interference cancellation. This is critical for practical ISAC systems where residual interference from co-existing waveforms can significantly degrade communication performance.

\item
We demonstrate the feasibility and performance of the proposed scheme through both analysis and simulations. We conduct an extensive simulation study comparing our proposed scheme with (1) a traditional OFDM system with embedded pilot subcarriers, and (2) the method in reference \cite{Wang2024-8}. Previous works often evaluated only sensing or communication separately. Our comparative analysis offers joint evaluation of both sensing and communication performance, including Bistatic Range Doppler accuracy, channel estimation accuracy, and BER. The results demonstrate that our method achieves robust gains across all three metrics, even under realistic channel conditions and imperfect sensing.
\end{itemize}

Compared to conventional OFDM systems with embedded pilot subcarriers, the
advantage of using a superposed FMCW signal as a pilot signal is summarized
as follows.

1) In conventional OFDM systems, the channel estimation is based on the embedded pilot subcarriers, which needs interpolation to generate the channels at non-pilot OFDM symbols and non-pilot subcarrers, which results
in inaccurate channel estimation in fast time-variant channel. Using
superposed FMCW signal to do channel estimation can track the fast
time-variant channel and estimate the time-variant channel more
accurately. Simulation results indicate that in a rapidly time-varying
channel (with high Doppler frequency), the proposed scheme delivers
superior sensing performance, achieves more accurate channel estimation,
and offers better BER performance compared to the conventional OFDM
systems.

2) Due to the periodic nature of OFDM symbols with embedded pilot
subcarriers within the OFDM frame, ghost targets can appear in the radar
range-Doppler map (RDM). The presence of these ghost targets compromises
the ability to detect weak targets, while our proposed method has
advantage in detecting weak targets. Simulation results show that, 
in terms of detecting weak targets, using the FMCW signal for sensing
surpass decision feedback-based sensing and channel estimation-based
sensing in conventional OFDM systems. Moreover, the Co-FMCW-OFDM yields superior sensing performance without incurring the
associated delays inherent in decision feedback mechanisms.

3) The additional advantage of Co-FMCW-OFDM is that: the superposed FMCW signal can be always there, like a broadcasting signal. Receivers can use it for
communication and/or sensing. Sometimes, the receiver does not need
communication, or when the SNR is too low for communication, the receiver
can still use the signal to sense the environment, which can provide real-time information to enable some critical applications.

The remainder of the paper is structured as follows. Section~\ref{system-model0} presents the proposed system model. Section~\ref{bistatic sensing} details FCCR sensing and DMD sensing with a superposed FMCW signal. Section~\ref{sensing-aided} presents the sensing-aided channel estimation and FMCW interference cancellation algorithms. Section~\ref{performance-evaluations} provides performance evaluations for the Co-FMCW-OFDM. Section~\ref{performance-comparison} first compares the sensing and data demodulation performance of the Co-FMCW-OFDM scheme with that of conventional OFDM systems employing embedded pilot subcarriers. It then compares the Co-FMCW-OFDM scheme with a conventional OFDM-plus-FMCW system as described in  \cite{Wang2024-8}. Our work is concluded in Section~\ref{conclusion}.

\section{The coordinated FMCW and OFDM system} \label{system-model0}

In this section, we present the system model, the transmitted signal model, and the received signal model.

\subsection{System Model} \label{system-model}

\begin{figure}[!]
	\centering
	\includegraphics[width=\linewidth]{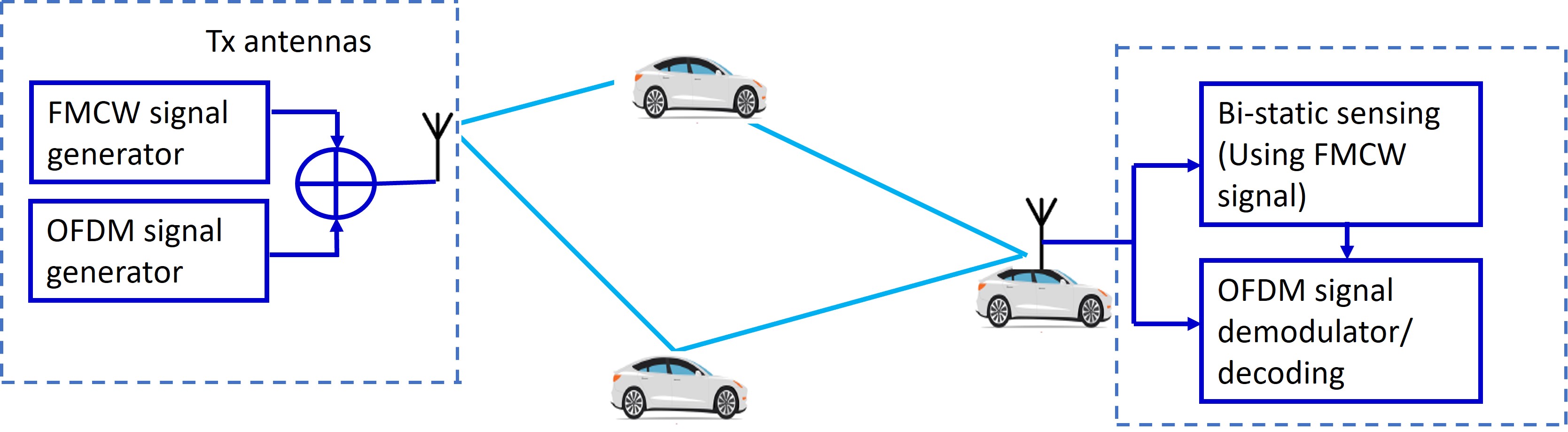}
\caption{Coordinated FMCW and OFDM scheme for ISAC}
	\label{fig1}
\end{figure}

The proposed Co-FMCW-OFDM design is shown in Fig.~\ref{fig1}. In the
coordinated system, a communication transceiver shares the same
transmitter antenna as the FMCW radar. The FMCW radar and communications
transceiver are coordinated and synchronized.

  The same model can be adapted as an ISAC system with superposed FMCW
  and OFDM waveform, where the FMCW signal is designed as a pilot signal
  to enable the channel estimation and bistaic sensing at the receiver.

  \subsection{Transmitted Signal Model}\label{transmitted-signal-model}

\begin{figure}[!]
	\centering
	\includegraphics[width=\linewidth]{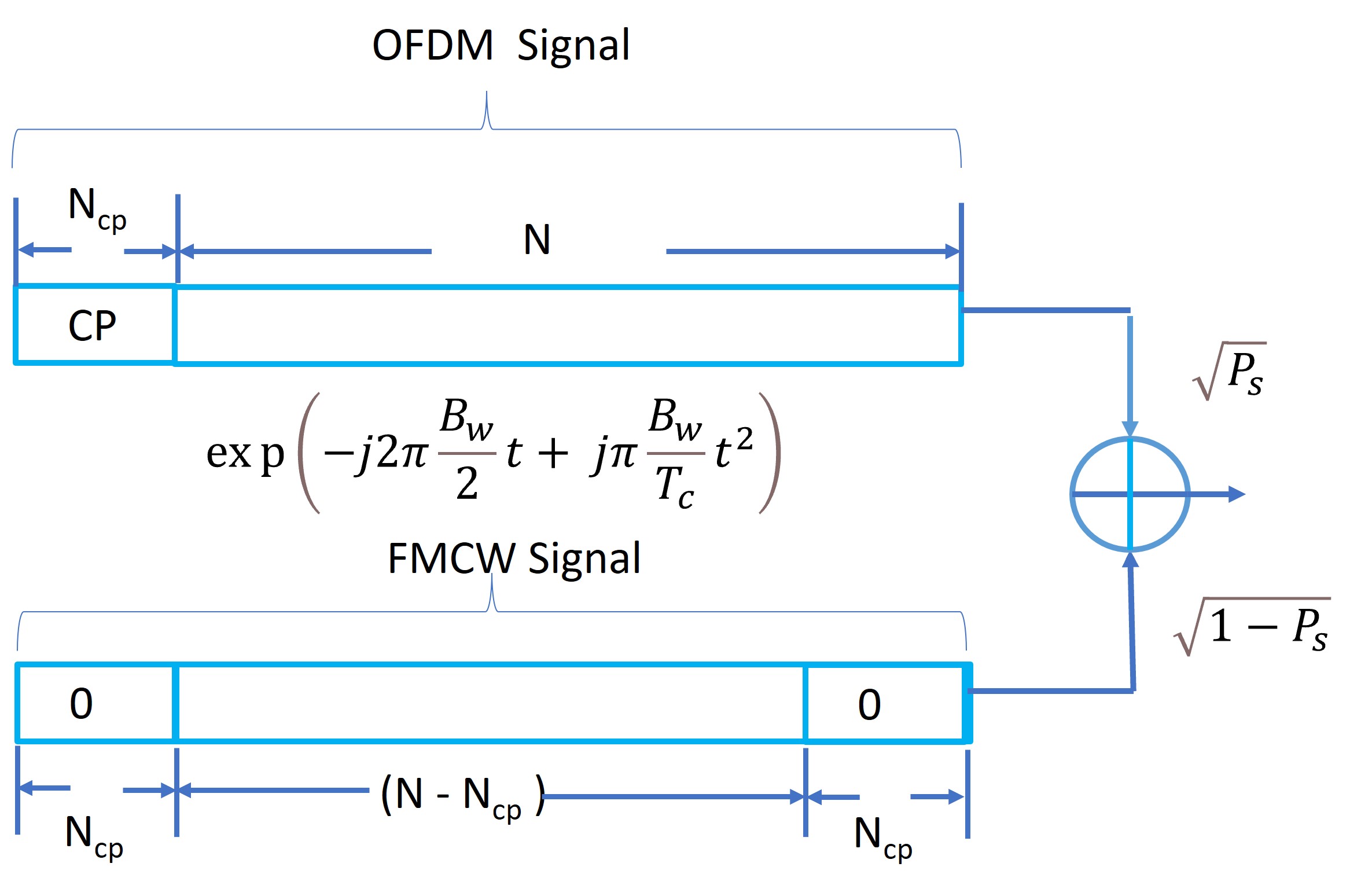}
\caption{One symbol of Co-FMCW-OFDM waveform}
	\label{fig2}
\end{figure}

 Fig.~\ref{fig2} illustrates one symbol of the proposed Co-FMCW-OFDM signal for ISAC. The radar transmits the FMCW signal known to the communication receiver.  The OFDM communication signal can be written as
\begin{equation}
s_{c}(t) = \sum_{m = 0}^{M - 1}{\sum_{n = 0}^{N - 1}{a_{m,n}\ e^{j2\pi n\mathrm{\Delta}ft}}}\Pi\left( t - mT_{sym} \right),
\end{equation}
where, \(T_{sym}\) is the duration of one OFDM symbol,
\(T_{sym} = T + \ T_{cp}\), \(\mathrm{\Delta}f = \frac{1}{T}\) is
subcarrier spacing, \(T_{cp}\) is duration of Cyclic Prefix (CP),
\(a_{m,n}\) is the multiplexed communication data symbol on \(n\)th
subcarrier of \(m\)th OFDM symbol. \(\Pi(t)\) is a rectangular pulse of
width \(T_{sym}\), \(N\) is number of subcarriers, \(M\) is number
of OFDM symbols.

 Sampling the \(s_{c}(t)\) with sampling period   \(T_{s} = \frac{1}{N\Delta f}\), we will get   \(N_{a} = N + N_{cp}\) samples for each OFDM symbol, where \(N_{cp}\)   is the length of CP in number of samples.

The FMCW signal is used for sensing and channel estimation at the
communication receiver. The baseband FMCW signal is defined as follows,
\begin{equation}\label{eq-FMCW}
s_{r}(t) =\begin{cases}
0 & mT_{sym}\  \leq t \leq mT_{sym} + T_{cp} \\
\exp\left( \varphi(t) \right)& {mT_{sym} + T}_{cp} < t \leq mT_{sym} + T\\
0& mT_{sym} + T < t \leq {mT_{sym} + T}_{cp} + T 
\end{cases}
\end{equation}
where \(\varphi(t) = j\pi\frac{B_{w}}{T_{c}}t^{2} - j2\pi\frac{B_{w}}{2}t\), \(m = 0,\ 1,\ \cdots,\ M - 1,\)
\(B_{w}\) is the bandwidth, and \(T + T_{cp} = T_{sym}\) is the symbol
duration of one FMCW symbol (which is same as the one OFDM symbol
duration). \(T_{c} = T - T_{cp}\) is the actual non-zero duration of one
FMCW symbol. According to equation~(\ref{eq-FMCW}), there is an all-zero cyclic prefix in \(s_{r}(t)\). This enables us to use the FCCR algorithm \cite{Zeng2020-7}  for sensing, which will be discussed in the next sections.

To control the out of band emission (OOBE), we need to apply a pulse shaping filter to \(s_{c}(t)\) and \(s_{r}(t)\). After filtering, the transmitted Co-FMCW-OFDM signal can be written as,
\begin{eqnarray}
x(t) &=& \sqrt{{1 - P}_{s}}\int_{- \infty}^{\infty}{g_0(t - \tau)}s_{r}(\tau){\mathrm d}\tau \nonumber \\
&&
+ \sqrt{P_{s}}\int_{- \infty}^{\infty}{g_0(t - \tau)}s_{c}(\tau){\mathrm d}\tau,
\end{eqnarray}
where \(g_0(t)\) is the impulse response of the pulse shaping filter.
Assume that the total power budget is 1, and \(P_{s}\) and \(1 - P_{s}\) are
the power allocated to OFDM signal and FMCW signal, respectively. \(P_{s}\)
is a design parameter, which can be flexibly adjusted to balance the
sensing and communication performance.

\subsection{Received Signal Model} \label{received-signal-model}

At the communication receiver, the signal is filtered by the the same filter $g_0(t)$ (matchced filtering). After the filtering, the received signal can be written as
\begin{eqnarray}
y(t) &=& \sqrt{{1 - P}_{s}}\sum_{p = 1}^{P}h_{p}e^{j2\pi f_{d,p}t} \nonumber\\
&& \cdot  \int_{- \infty}^{\infty}{g\left( t - \tau_{p} - \tau \right)s_{r}(\tau)d\tau} \nonumber\\
&&  + \sqrt{P_{s}}\ \sum_{p = 1}^{P}h_{p}e^{j2\pi f_{d,p}t}\nonumber\\
&& \cdot \int_{- \infty}^{\infty}{g\left( t - \tau_{p} - \tau \right)s_{c}(\tau)d\tau} + \ \eta(t),
\end{eqnarray}
where $g=g_0*g_0$ (convolution), \(\eta(t)\) is the additive white Gaussian Noise (AWGN), \(P\) is the number of targets, \(h_{p}\) is the target reflection coefficient, \(\tau_{p}\) is the delay associated with the path reflected by target
\emph{p}, \(f_{d,p} = \frac{\vartheta_{p}}{c}f_{c}\) is the Doppler
frequency shift induced by the moving of target \(p\), \(f_{c}\) is
carrier frequency, \(\vartheta_{p}\) is the relative speed of the target
\(p\) and \(c\) is the speed of light. Radar sensing is to find the
delay \(\tau_{p}\) and Doppler frequency shift \(f_{d,p}\).

We consider a practical channel model, where the delay \(\tau_{p}\) may
not be an integer multiple of sampling period \(T_{s}\). Let
\(\tau_{p} = l_{p}T_{s} + \alpha_{p}\), with
\(0 \leq \alpha_{p} < T_{s}\) and \(l_{p}\) is an integer number. The raised cosine filter can be used as the pulse shaping filter, whose time domain impulse response is given as
\begin{equation}
g(t) = \frac{\sin(\pi t/T_s)}{\pi t/T_s} \cdot \frac{\cos(\pi \beta t/T_s)}{1-(2\beta t/T_s) ^2},
\end{equation}
where $\beta$ is the roll-off factor. In practical implementations, the filter is time-limited by truncating $g(t)$ to a finite duration. Specially, it is assumed that $g(t)$ is time-limited to the interval [$-\Delta T_s$, $\Delta T_s$ ]; that is  \(g(t)=0\) for \(t > \Delta T_{s}\) or \(t < - \Delta T_{s}\). Then the sampled received signal is
\begin{align}
y( nT_s) =& \sqrt{1 - P_s}\sum_{p = 1}^{P}\sum_{k = - \Delta}^{\Delta}h_{p,k} s_{r}( n - l_{p} - k )T_{s} ) \nonumber \\
&\cdot {\rm e}^{j2\pi f_{d,p}nT_s} \nonumber \\
&+ \sqrt{P_s}\sum_{p = 1}^{P}\sum_{k = - \Delta}^{\Delta}h_{p,k} s_{c} ( n - l_{p} - k )T_{s} ) \nonumber\\
&\cdot {\rm e}^{j 2\pi f_{d,p}nT_s}+\eta ( nT_s), \label{eq-chan-model}
\end{align}
where 
$h_{p,k} = h_{p} g\left( kT_{s} - \alpha_{p} \right), \ p = 1,\cdots P;\ k = - \mathrm{\Delta}, \cdots, \mathrm{\Delta}$.
Note that  since the pulse shaping filter is a raised cosine filter,
it satisfies the property that \(g\left( nT_{s} \right) = 1\) when
\(n = 0\) and \(g\left( nT_{s} \right) = 0\) when \(n \neq 0\).
Therefore, when the delays \(\tau_{p}\) of all paths are integer
multiples of the sampling period \(T_{s}\), i.e., \(\alpha_{p}\ \)= 0,
the effective channel will contain \(P\) paths. However, if \(\tau_{p}\)
is not an integer multiple of \(T_{s}\), each path will expand to
\((2\mathrm{\Delta} + 1)\) effective paths, with delay of each effective
path being
\(\left( l_{p} + k \right)T_{s},\ k = - \mathrm{\Delta},\ \cdots,\ \mathrm{\Delta}.\ \)
When delays \(\tau_{p}\) of all paths are not the integer multiples of
\(T_{s}\), totally there will be \((2\mathrm{\Delta} + 1)P\) effective
paths.

\begin{figure}[!]
	\centering
	\includegraphics[width=\linewidth]{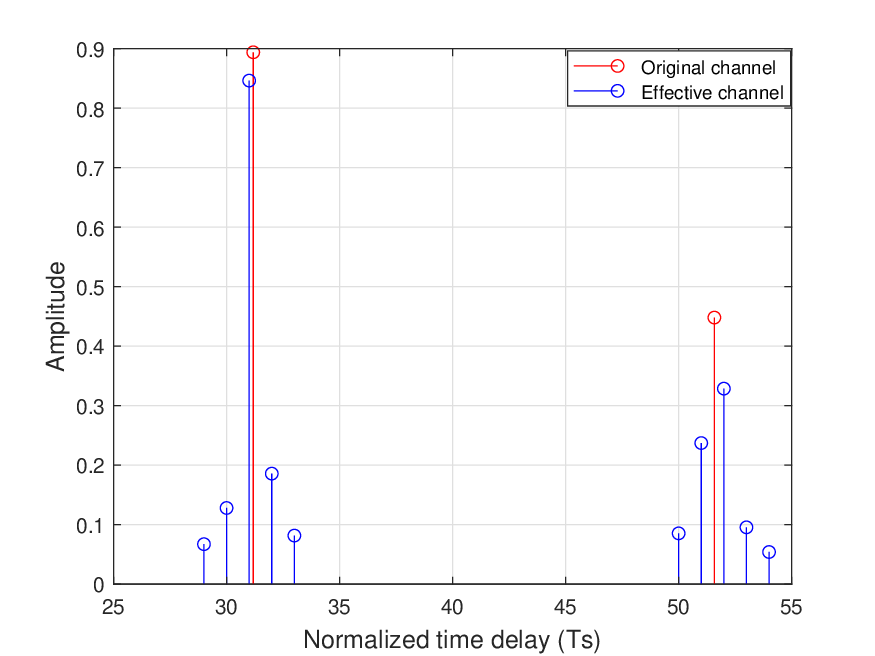}
\caption{An example of original and effective channel profile}
	\label{fig3}
\end{figure}

Fig.~\ref{fig3} illustrates an example of the original channel profile and the
corresponding effective channel profile. The original channel is a
two-path channel with delay \(\tau_{1}\) = 0.5075 \(\mu s\) and
\(\tau_{2}\) = 0.8395 \(\mu s\). Given a sampling period \(T_{s}\) =
0.016276 \(\mu s\), we have \(\tau_{1}\) = 31.18\(T_{s}\) and
\(\tau_{2}\) = 51.58\(T_{s}\). Assume \(\mathrm{\Delta}\  = 2\), the
effective channel will consist of 10 paths, with path delay being [29, 30, 31, 32, 33, 50, 51, 52, 53, 54]\(T_{s}\).

In general, given an original channel with \(P\) paths, the effective
channel will consist of \((2\Delta + 1)P\) paths, whose channel
coefficient, delay and Doppler frequency can be written as follows:
\begin{equation}\label{eq-effect-chan}
\left( {\widetilde{h}}_{l},\ {\widetilde{\tau}}_{l},\ {\widetilde{f}}_{d,l} \right),\ l = 1,\cdots,\widetilde{P},
\end{equation}
where \(\widetilde{P} = (2\mathrm{\Delta} + 1)P\) is the number of
effective paths,
\({\widetilde{\tau}}_{l} = \left( l_{p} + k \right)T_{s}\) is the time
delay, \({\widetilde{f}}_{d,l} = f_{d,p}\) is the Doppler frequency,
\({\widetilde{h}}_{l} =\ h_{p} g\left( kT_{s} - \alpha_{p} \right)\) is the channel
coefficient,
\(l = (2\mathrm{\Delta} + 1)(p - 1) + \mathrm{\Delta} + k + 1\),
\(p = 1,\cdots,P;k = - \mathrm{\Delta},\cdots,\mathrm{\Delta}\ .\)

In the following we will use the effective channel model consisting of
\(\widetilde{P}\ \)paths, with the channel coefficient, delay and
Doppler frequency defined by equation~(\ref{eq-effect-chan}). For notation simplicity, we will drop the sampling period $T_s$ in all the equations, 
that is, we will use $y(n)$ for \(y\left( nT_{s} \right)\), and similarly for others.

The signal \(y(n)\) is composed of \(M\) symbols
with each symbol having a length of \(N_{a}\) samples.  In each symbol,
the first \(N_{cp}\) samples are discarded. After discarding CP, the
symbol \(m\) is denoted as
\({\overline{y}}_{m}\left( n\right) (n = 0,\ 1,\ \cdots,\ N - 1,\ m = 0,\ 1,\ \cdots,\ M - 1).\)
Assume that the normalized delay is always smaller than the CP length,
i.e., \((l_{p} + \mathrm{\Delta}) < N_{cp}\), then the received signal
for symbol \(m\) can be expressed as,
\begin{equation}
{\overline{y}}_{m}\left( n\right) = \sum_{l = 1}^{\widetilde{P}}{{{\widetilde{\widetilde{h}}}_{l}s}_{m}(\left\langle n - {\widetilde{\varepsilon}}_{l} \right\rangle_{N}}) {\rm e}^{j2\pi nv_{l}}{\rm e}^{j2\pi mN_{a}v_{l}} + \eta_{m}\left( n\right),
\end{equation}
where \(v_{l} = {\widetilde{f}}_{d,l}T_{s}\) is normalized Doppler frequency, \({\widetilde{\varepsilon}}_{l} = {\widetilde{\tau}}_{l}/T_{s}\)
is normalized delay, \({\widetilde{\widetilde{h}}}_{l} = {\widetilde{h}}_{l}e^{j2\pi N_{cp}v_{l}}\),
\(\eta_{m}\left( n\right)\) is the additive white Gaussian noise
(AWGN).
\(s_{m}\)(\(n)\)=\( \sqrt{{1 - P}_{s}}s_{r,m}\left( \ n\  \right)  + \sqrt{P_{s}}s_{c,m}\left( \ n\  \right)\),
\(s_{r,m}\left( n\right)\ \)and \(s_{c,m}\left( n\right)\)
are samples of the FMCW signal and the OFDM signal for symbol \(m\),
respectively. \(\left\langle n \right\rangle_{K} \triangleq n\mod K\)
denote the remainder of $n$ modulo $K$.

\section{Bistatic Sensing with the Superposed FMCW signal} \label{bistatic sensing}

\begin{figure}[!]
	\centering
	\includegraphics[width=\linewidth]{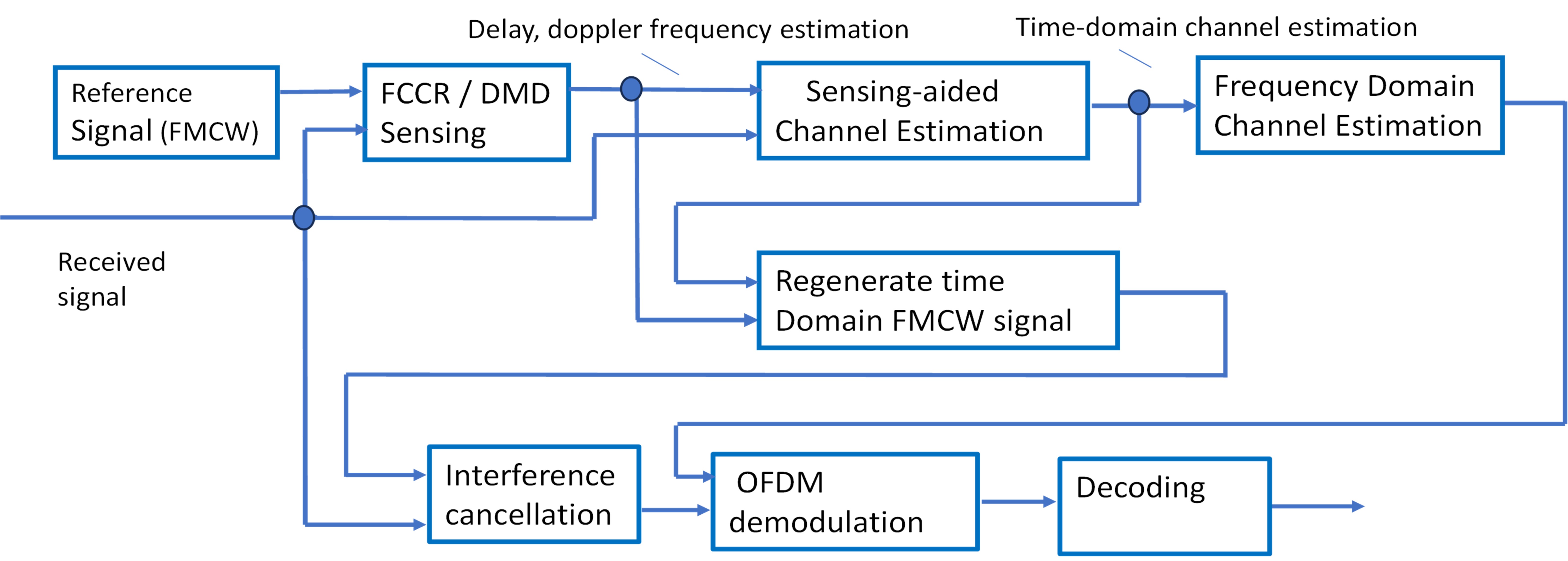}
\caption{Receiver Algorithm Diagram}
	\label{fig4}
\end{figure}

In this section, we present bistatic sensing algorithms with the superposed FMCW signal. Fig.~\ref{fig4} shows the diagram of the proposed receiver algorithm. Both FCCR \cite{Zeng2020-7} and the DMD sensing algorithms can be used to estimate the delay and Doppler frequency. By utilizing the information obtained from sensing and channel estimation, interference from FMCW signals is reconstructed and eliminated from the received signal in the time domain before OFDM demodulation.
 
\subsection{FCCR Sensing Algorithm} \label{fccr-sensing-algorithm}
The FCCR sensing algorithm \cite{Zeng2020-7} was derived for monostatic radar sensing based on the maximum likelihood (ML) principle and can be used for any waveform with a CP structure. Theoretically, it is nearly optimal given that the delay is within the CP range. Here we consider using the FCCR for bistatic sensing with the superposed FMCW signal. 

In FCCR, the cyclic correlation of the received signal with the transmitted signal is used for range and speed detection. For bistatic sensing in Co-FMCW-OFDM, we only know the transmitted FMCW signal at the receiver. Thus, we can only use the cyclic correlation of the received signal with the FMCW signal.  After discarding the CP, we denote the received signal for \(m\)-th symbol duration as
\({\overline{y}}_{m}\left( n\right)\), and the transmitted FMCW signal for
\(m\)-th symbol duration as
\(s_{r,m}\left( n\right) \), \( n = 0,\ 1,\ \cdots,\ N - 1\). The
cyclic correlation between the received signal
\({\overline{y}}_{m}\left( n\right)\) and the transmitted FMCW signal
\(s_{r,m}\left( n\right)\) is
\begin{equation}
r(n,m) = \sum_{l = 0}^{N - 1}{{\overline{y}}_{m}\left( l\right)s_{r,m}^{*}\left( \left\langle l - n \right\rangle_{N}\right)},
\end{equation}
$$m=0,1,\cdots,M-1.$$
With similar derivations as in \cite{Zeng2020-7}, we can get the range and speed detection with the cyclic correlation $r(n,m)$. To save space, we omit the details here. The cyclic correlation $r(n,m)$ can be computed by the fast Fourier transform (FFT) as shown in \cite{Zeng2020-7}. After getting $r(n,m)$, we compute the FFT of \(r(n,m)\) to get \(R(n,k)\) as
\begin{equation}\label{RDM}
R(n,k) = \sum_{m = 0}^{M - 1}{r(n,m)}e^{j\frac{- 2\pi mk}{M_1}},
\end{equation}
where $M_1 \ge M$ is a positive integer for the FFT size. Selecting $M_1 >M$ can improve the accuracy of Doppler frequency estimation. Based on the range-Doppler map (RDM) \(\left| R(n,k) \right|\), we can find the delay and Doppler frequency of the \(P\) targets \cite{Zeng2020-7}.

\subsection{DMD Sensing Algorithm} \label{dmd-sensing-algorithm}
In a conventional FMCW radar receiver, the received signal is mixed with the
reference signal in the analog domain \cite{Richards2022-10,Venon2022-11}. This mixed signal is then passed through a low-pass filter to remove high-frequency
components and isolate the beat frequency,  to reveal the target's range and speed. However, this FMCW radar processing cannot be used in a common communication receiver for bistatic sensing, as a common communication receiver does not have the transmitted analogue FMCW signal and the required hardware. Thus, we propose the DMD sensing algorithm to estimate the delay
and Doppler frequency. The key differences between the conventional FMCW
sensing algorithm and the DMD sensing algorithm are as follows: 1) In
the DMD sensing algorithm, mixing is performed in the digital domain; 2)
As the beat frequency is relatively low, a downsampler is used in
the DMD algorithm to reduce the sample size, thereby decreasing the implementation complexity.

Adopting the effective channel model described in (\ref{eq-chan-model}), we can write \(y\left( n\right)\) as,
\begin{eqnarray}
y\left( n\right) = \sqrt{{1 - P}_{s}}\sum_{l = 1}^{\widetilde{P}}{{{\widetilde{h}}_{l}\ s}_{r}\left(  n - {\widetilde{\varepsilon}}_{l} \right)}{\ e}^{j2\pi{\widetilde{f}}_{d,l}nT_{s}} \nonumber\\
+  \sqrt{P_{s}}{\sum_{l = 1}^{\widetilde{P}}{\ {\widetilde{h}}_{l}s}_{c}\left(  n - {\widetilde{\varepsilon}}_{l} \right)}e^{j2\pi{\widetilde{f}}_{ d,l }nT_{s}} +  \eta\left( n\right).
\end{eqnarray}
The known sampled FMCW signal can be written as,
\begin{equation}
s_{ref}\left( n\right) = s_{r}\left( n\right) = {\rm exp}\left( j\pi\frac{B_{w}}{T_{c}}{(nT_{s})}^{2} - j2\pi\frac{B_{w}}{2}nT_{s} \right).
\end{equation}
Multiplying the \(s_{ref}\left( n\right)\) with the conjugate of
received signal \(y\left( n\right)\), we get,
\begin{align}
&\Upsilon\left( n\right)= \ s_{ref}\left( n\right)y^{*}\left( n\right)  \nonumber\\
&= \sqrt{\left( 1 - P_{s} \right)}\sum_{l = 1}^{\widetilde{P}}{{{\widetilde{h}}^{*}_{l}s}_{r}\left( n\right)}s_{r}^{*}\left(  n - {\widetilde{\varepsilon}}_{l} \right){\rm e}^{- j2\pi{\widetilde{f}}_{d,l}nT_{s}}  \nonumber\\
&+ \sqrt{P_{s}}\sum_{l = 1}^{\widetilde{P}}{{\widetilde{h}}_{l}^{*}s_{r}\left( n\right)s_{c}^{*}(n - {\widetilde{\varepsilon}}_{l}}) {\rm e}^{- j2\pi{\widetilde{f}}_{d,l}nT_{s}} + s_{r}\left( n\right)\eta^{*}\left( n\right)  \nonumber\\
&= \sqrt{\left( 1 - P_{s} \right)}\sum_{l = 1}^{\widetilde{P}}{\widetilde{h}}_{l}^{*}{\rm e}^{j2\pi f_{bt}nT_{s}}{\rm e}^{- j\pi\frac{B_{w}}{T_{c}}\left( {\widetilde{\tau}}_{l}^{2} +T_{c}{\widetilde{\tau}}_{l} \right)}{\rm e}^{- j2\pi{\widetilde{f}}_{d,l}nT_{s}} \nonumber\\
&+\sqrt{P_{s}}\sum_{l = 1}^{\widetilde{P}}{{\widetilde{h}}_{l}^{*}s_{r}(n)s_{c}^{*}(n - {\widetilde{\varepsilon}}_{l}}){\rm e}^{- j2\pi{\widetilde{f}}_{d,l}nT_{s}}+s_{r}(n)\eta^{*}(n), \label{eq-DMD}
\end{align}
where \(f_{bt} = \frac{B_{w}}{T_{c}}{\widetilde{\tau}}_{l}\) is the
beat frequency. In (\ref{eq-DMD}), the first term is the desired signal, the
second term is the interference from the OFDM signal, and the third term
is the noise. Note that the first term is a narrowband signal; we can use
a low-pass filter to reduce the impact of interference and noise. Denote the output of the low-pass filter as \(\Lambda\left( n\right)\). The relationship
between \(\Lambda\left( n\right)\) and \(\Upsilon\left( n\right)\) can be written as,
\begin{equation}
\Lambda \left( n\right) = \sum_{l = 1}^{N_{F}}\lambda_{l}\Upsilon\left( (n - l)\right),
\end{equation}
where \(N_{F}\) is the number of taps of the low-pass filter, and
\(\left\{ \lambda_{l},\ l = 1,\ldots N_{F} \right\}\) is the coefficient
of the low-pass filter.

\(\Lambda\left( n\right)\) can be down sampled by a factor of
\(D\) to reduce the sample rate and implementation complexity. The choice of \(D\) shall fulfill 1)
\(1/{(DT}_{s}) \geq 2f_{bt}\), 2) \(N_{a}\) and \(N_{cp}\) are divisible
by \(D\). After down sampling, an FMCW symbol contains \(N_{T} = N_{a}/D\)
samples including the gap, and \(N_{D} = (N - 2N_{cp})/D\) samples excluding
the gap. Denote \(z(n,m)\) \((n = 0,1,\cdots,N_{D} - 1\),
\(m = 0,\ 1,\ \cdots,\ M - 1\)) as the \(n\)-th sample of the
\(m\)-th FMCW symbol. The relationship between \(z(n,m)\) and
\(\Lambda\left( n\right)\) is
\begin{equation}
z(n,m) = \Lambda\left( (n + N_{b} + {mN}_{T})D\right),
\end{equation}
$$ n = 0,1,\cdots,N_{D} - 1, m = 0,\ 1, \cdots, M - 1$$
where, \(N_{b} = N_{cp}/D\) is the number of samples to skip for each
FMCW symbol during the down sample process.

To get the RDM, we first calculate the column wise FFT of
\(z(n,m)\), which can be written as,
\begin{equation}
U(l,m) = \sum_{n = 0}^{N_{D} - 1}{z(n,m)}e^{- j\frac{2\pi nl}{N_{D}}},
\end{equation}
$$ l = 0,1,\cdots,N_{D} - 1, m = 0, 1, \cdots, M - 1.$$
Then we calculate the row-wise IFFT of \(U(l,m)\), which can be written
as,
\begin{equation}
\overline{R}(l,k) = \sum_{m = 0}^{M  - 1}{U(l,m)}e^{j\frac{2\pi mk}{M_1}},
\end{equation}
$$ l = 0,1,\cdots,N_{D} - 1, k = 0, 1, \cdots, M_1 - 1.$$
\(\overline{R}(k,l)\) represents the sensing RDM. Based on
\(\left| \overline{R}(k,l) \right|\), we can determine which points are
targets.

\section{Sensing-aided channel estimation and interference cancellation} \label{sensing-aided}
In this section, we present a sensing-aided channel estimation, which estimates the path delays, the Doppler frequencies, and the channel coefficients of the effective channel based on the delay and Doppler frequency estimation obtained from the sensing procedure. By using the information
obtained from sensing and channel estimation, interference from FMCW
signals is reconstructed and eliminated from the received signal in the
time domain before OFDM demodulation.

\subsection{Path delay and Doppler frequency of the effective channel} \label{sensing-aided-ch}
At the sensing stage, we have found the normalized delay and Doppler frequency of the $P$ targets: \(\left( {\widehat{l}}_{p},\ {\widehat{k}}_{p} \right)\) $(p = 1, \cdots,P)$. Due to the pulse shaping filtering, the effective channel has \(\widetilde{P} = (2\Delta + 1)P\) paths (see equation (\ref{eq-chan-model})). The pulse shaping creates $2\Delta+1$ pseudo-targets (include the real target) around each real target. For channel estimation purposes, at each of the detected targets, we need to estimate all the \((2\Delta + 1)\) paths. Fortunately, based on the sensing results, we have the estimations of the normalized delay and Doppler frequency of the pseudo-targets as:  \(\left\{ \left( {\widehat{l}}_{p} - \Delta, {\widehat{k}}_{p} \right),\ \left( {\widehat{l}}_{p} - \Delta+ 1\ ,{\widehat{k}}_{p} \right),\ \ldots,\ ({\widehat{l}}_{p} + \Delta, {\widehat{k}}_{p}) \right\}, \) \(\ (p = 1,\cdots,P\), $ j=-\Delta, \cdots, \Delta) $ The strength of each pseudo-target can also be computed directly from the RDM (see equation (\ref{RDM})) as $ A_{p,j}= |R(D_{p,j},K_{p,j})|$, where $ D_{p,j} = \widehat{l}_{p} + j$ and $ K_{p,j} = \widehat{k}_{p}$ are normalized delay and Doppler frequency of the $ (2\Delta + 1)$ pseduo-targets within the target cluster $p$. Thus, after the sensing, we have the path delays and Dopplers: \({\widehat{\tau}}_{p}\) and \({\widehat{f}}_{d,p}\),  for all the \(\widetilde{P}\) paths of the effective channel.

\subsection{Sensing-Aided Channel Estimation} \label{sensing-aided-ce}
After finding the path delays and Doppler frequencies, the remaining task is to estimate the channel coefficients of all the paths. We use a SIC algorithm to estimate the time domain channel coefficients. Firstly, we estimate the channel coefficient of the most significant path. Then, we regenerate the signal corresponding to the most significant path and cancel it from the received signal before we estimate the channel coefficient of the second significant path. The procedure continues until we estimate the channel coefficients for all the $\widetilde{P}$ paths. For simplicity and without loss of generality, in the following we assume that the paths are ordered according to their strengths. Let $ \widetilde{h}_p$, $ \widetilde{\tau}_p$ and $ \widetilde{f}_{d,p}$ ($ p = 1,\cdots,\widetilde{P})$ denote the time domain channel coefficient, delay and Doppler frequency of the $p$th significant path, respectively. Denote $\widetilde{\epsilon}_p$ as the normalized delay of the $p$th significant path, with $\widehat{l}_p$ being its estimation, and $\widehat{f}_{d,p}$ as the Doppler frequency estimation for the $p$th significant path. Consequently, the reference signal for the $p$th significant path can be regenerated as,
\begin{equation}\label{eq-ref-signal}
{\widetilde{r}}_{p}\left( n\right) = \sqrt{{1 - P}_{s}}s_{r}\left( n- {\widehat{l}}_{p} \right)e^{j2\pi{\widehat{f}}_{d,p}nT_{s}}\ ,
\end{equation}
$$ p = 1,\cdots,\widetilde{P}.$$

The estimation of the channel coefficient for the most significant path $(p = 1)$ can be written as
\begin{equation}\label{eq-first-path}
{\widehat{h}}_{1} = \frac{\ \ \sum_{n = 0}^{MN_{a} - 1}{y\left( n\right)\left( {\widetilde{r}}_{1}\left( n\right) \right)^{*}}}{\sum_{n = 0}^{MN_{a} - 1} | {\widetilde{r}}_{1}\left( n\right) |^{2}}.
\end{equation}
After $\widehat{h}_{1}$ is obtained, the signal corresponding to the most significant path can be regenerated as
\begin{equation} \label{eq-re-gen-path}
{\widetilde{y}}_{1}\left( n\right) = \widehat{h}_1\sqrt{{1 - P}_{s}}s_{r}\left( n- {\widehat{l}}_{p} \right)e^{j2\pi{\widehat{f}}_{d,p}nT_{s}}.
\end{equation}
Assume that we have obtained the estimation of the channel coefficient for the $p$th significant path $(p<\widetilde{P})$.  The estimation of the channel estimatin coefficient for the $(p+1)$th path can be written as
\begin{equation} \label{eq-ch-pluse-1}
{\widehat{h}}_{p+1} = \frac{\ \ \sum_{n = 0}^{MN_{a} - 1}{(y( n)-\widetilde{y}_p(n))\left( {\widetilde{r}}_{p+1}\left( n\right) \right)^{*}}}{\sum_{n = 0}^{MN_{a} - 1} | {\widetilde{r}}_{p+1}\left( n\right) |^{2}},
\end{equation}
where $\widetilde{y}_p(n)$ can be written as
\begin{equation} \label{eq-yp-n}
{\widetilde{y}}_{p}\left( n\right) = \sum_{i=1}^{p}{ \widehat{h}_i\sqrt{{1 - P}_{s}}s_{r}\left( n- {\widehat{l}}_{i} \right)e^{j2\pi{\widehat{f}}_{d,i}nT_{s}}}.
\end{equation}

According to \cite{zeng2018-9,Gaudio2019-12}, when the Doppler frequency \(f_{d,p} \ll \Delta f\), the frequency domain channel estimation
\(\widehat{H}(k,m)\ \) at OFDM symbol \(m\) and subcarrier \(k\) can be generated from the estimated time domain channel coefficients, delays, and Doppler frequencies as
\begin{equation}
\widehat{H}(k,m) = \sum_{p = 1}^{\widetilde{P}}{{\widehat{h}}_{p}e^{- j2\pi k\mathrm{\Delta}f{\widehat{\tau}}_{p}}e^{j2\pi mT_{sym}{\widehat{f}}_{d,p}}}.
\end{equation}
Similarly, the actual frequency domain channel $H(k,m)$ at OFDM symbol $m$ and subcarrier $k$ can be written as
\begin{equation}
{H}(k,m) = \sum_{p = 1}^{\widetilde{P}}{{\widetilde{h}}_{p}e^{- j2\pi k\mathrm{\Delta}f{\widetilde{\tau}}_{p}}e^{j2\pi mT_{sym}{\widetilde{f}}_{d,p}}}.
\end{equation}
The normalized mean square error (NMSE) of the frequency domain channel estimation is defined as
\begin{equation}
\delta_{H,NMSE} = {\mathbb{E}\left\{ \frac{\sum_{m = 0}^{M - 1}{\sum_{n = 0}^{N_{sc} - 1}} \left| \widehat{H}(k,m) - H(k,m) \right|^{2}} {\sum_{m = 0}^{M - 1}{\sum_{n = 0}^{N_{sc} - 1}} \left| H(k,m)  \right|^{2}} \right\}}.
\end{equation}
Generally speaking, $\delta_{H,NMSE}$ consists of two parts: 1) error caused by sensing, i.e., the difference between \({\widehat{f}}_{d,p}\) and
\({\widetilde{f}}_{d,p}\), and difference between
\({\widehat{\tau}}_{p}\) and \({\widetilde{\tau}}_{p}\); 2) error caused
by channel estimation, i.e., the difference between
\({\widehat{h}}_{p}\) and \({\widetilde{h}}_{p}\).

\subsection{Analysis on the Sensing-Aided Channel Estimation} \label{analysis}
In the following, we analyze the estimation error of the channel coefficient for the most significant path. Note that, as the FMCW signal is superimposed on the OFDM signal, the OFDM signal is treated as interference during the channel estimation process.
Assume that the estimations of \(\tau_{p}\) and \(f_{d,p}\) are ideal,
i.e., \({\widehat{l}}_{1} = {\widetilde{\epsilon}}_{1}\) and
\({\widehat{f}}_{d,1} = {\widetilde{f}}_{d,1}\). Substituting (\ref{eq-ref-signal}) into
(\ref{eq-first-path}), we get,
\begin{equation}
{\widehat{h}}_{1} = \ {\widetilde{h}}_{1} + \frac{I_{F,1} + I_{O,1} + I_{A,1}}{\sum_{n = 0}^{MN_{a} - 1}{\left( 1 - P_{s} \right)\left| s_{r}\left( n- {\widetilde{\epsilon}}_{1} \right) \right|^{2}\ }},
\end{equation}
where \(I_{F,1}\) denotes the interference from other paths of the FMCW
signal, and \(I_{O,1}\) denotes the interference from the OFDM signal and 
\(I_{A,1}\) denotes the interference from the AWGN.
\begin{eqnarray}
I_{F,1} = \sum_{n = 0}^{MN_{a} - 1}{\sum_{q = 2}^{\widetilde{P}}{\widetilde{h}}_{q}\left( 1 - P_{s} \right)s_{r}\left( n- {\widetilde{\epsilon}}_{q} \right)}s_{r}^{*}\left( n- {\widetilde{\epsilon}}_{1} \right), \\
I_{O,1} = \sum_{n = 0}^{MN_{a} - 1}{\sum_{q = 1}^{\widetilde{P}}{\widetilde{h}}_{q}\sqrt{P_{s}(1 - P_{s})}s_{c}\left( n- {\widetilde{\epsilon}}_{q} \right)}s_{r}^{*}\left( n- {\widetilde{\epsilon}}_{1} \right),\\
I_{A,1} = \sum_{n = 0}^{MN_{a} - 1} \sqrt{1 - P_{s}}e^{- j2\pi{\widetilde{f}}_{d,p}nT_{s}} s_{r}^{*} \left( n- {\widetilde{\epsilon}}_{1} \right)\eta\left( n\right) .
\end{eqnarray}
Assume that OFDM signal and FMCW signal are not correlated and
\(\mathbb{E}\left\{ \left| s_{r}\left( n\right) \right|^{2} \right\}\mathbb{= E}\left\{ \left| s_{c}\left( n\right) \right|^{2} \right\} = 1\).
The root mean square error (RMSE) of time domain channel estimation for the most significant path can be written as,
\begin{eqnarray}
\sigma_{h_{1}}& =& \sqrt{\mathbb{E}\left\{ \left| {\widehat{h}}_{1} - {\widetilde{h}}_{1} \right|^{2} \right\}}\nonumber \\
&=& \ \sqrt{\mathbb{\ E}\left\{ \left| \frac{I_{F,1} + I_{O,1} + I_{A,1}}{\sum_{n = 0}^{MN_{a} - 1}{\left( 1 - P_{s} \right)\left| s_{r}\left( n- {\widetilde{\epsilon}}_{1} \right) \right|^{2}\ }} \right|^{2} \right\}}\nonumber \\
&=& \sqrt{\frac{\mathbb{E}\left\{ \left| I_{F,1} \right|^{2} \right\}\mathbb{+ E}\left\{ \left| I_{O,1} \right|^{2} \right\}\mathbb{+ E}\left\{ \left| I_{A,1} \right|^{2} \right\} } {\mathbb{E}\left\{ \left| \sum_{n = 0}^{MN_{a} - 1}{\left( 1 - P_{s} \right)\left| s_{r}\left( n- {\widetilde{\epsilon}}_{1} \right) \right|^{2}\ } \right|^{2} \right\}}}.
\end{eqnarray}
When there is only one target and path delay associated with the path
reflected by the target is integer multiples of \(T_{s}\), i.e.,
\(P = \widetilde{P} = 1\), \(\sigma_{h_{1}}\) can be approximated by
\begin{equation}\label{eq-RMSE}
\sigma_{h_{1}} \approx \sqrt{\frac{T + T_{cp}}{T - T_{cp}}\frac{N}{N_{sc}}\frac{P_{s}}{\left( 1 - P_{s} \right)MN_{a}} + \frac{\sigma_{n}^{2}}{\left( 1 - P_{s} \right)MN_{a}}},
\end{equation}
where \(N_{sc} < N\) is the actual number of subcarriers to carry data
symbols and \(\sigma_{n}^{2}\) is the noise variance. Equation (\ref{eq-RMSE})
indicates that \(\sigma_{h_{1}}\) is inversly proportional to
\(\sqrt{MN_{a}}\).

\begin{figure}[!]
	\centering
	\includegraphics[width=\linewidth]{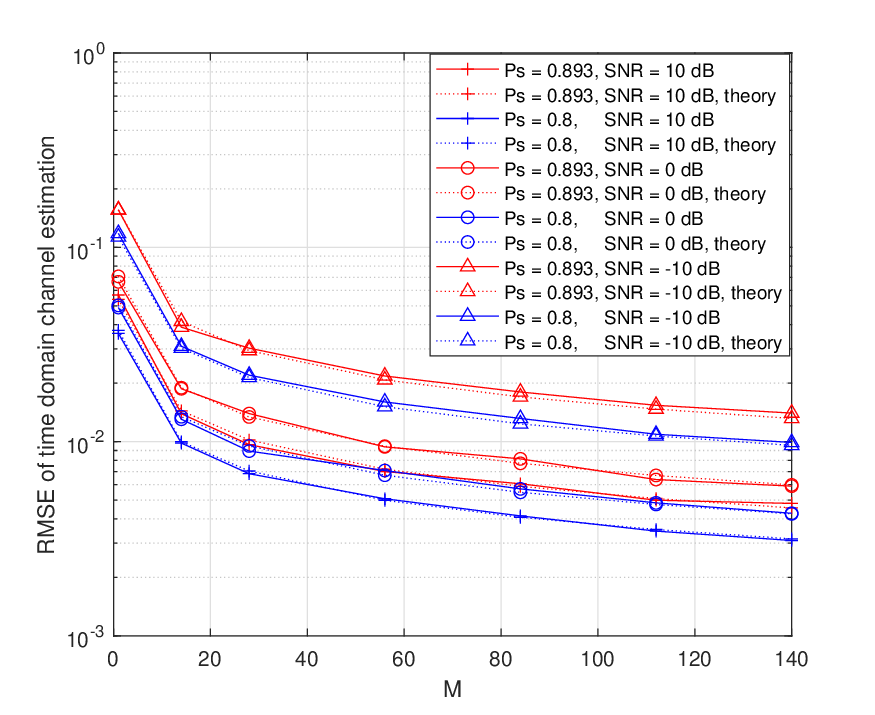}
\caption{Simulated and theoretic RMSE of time domain channel estimation}
	\label{fig5}
\end{figure}

Fig.~\ref{fig5} shows the simulated \(\sigma_{h_{1}}\ \) and theoretic RMSE
when there is one target and the path delay is an integer multiple of \(T_{s}\) (
\(P = \widetilde{P} = 1\)). The Doppler frequency is 900 Hz and
subcarrier interval is 15 KHz. From Fig.~\ref{fig5} we can see that
\(\sigma_{h_{1}}\ \) decreases as \(M\) and SNR increase
and the simulated \(\sigma_{h_{1}}\ \)and theortic \(\sigma_{h_{1}}\ \) are well aligned.

\subsection{Time domain interference cancellation} \label{time-domain-interference-cancellation}

In time domain, the regenerated FMCW signal can be written as,
\begin{equation}
{\widehat{r}}_{fmcw}\left( n\right) = \sum_{p = 1}^{\widetilde{P}}{\widehat{h}}_{p}\sqrt{{1 - P}_{s}}s_{r}\left( n- {\widehat{l}}_{p} \right)e^{j2\pi{\widehat{f}}_{d,p}nT_s}.
\end{equation}
The time domain interference cancellation can be written as 
\begin{align}\label{eq-SIC}
&y_{I}\left( n\right) = y\left( n\right) - {\widehat{r}}_{fmcw}\left( n\right) \nonumber \\
&= \sqrt{{1 - P}_{s}}\sum_{p = 1}^{\widetilde{P}}{{{\widetilde{h}}_{p}s}_{r}(n- {\widetilde{\epsilon}}_{p}})e^{j2\pi{\widetilde{f}}_{d,p}nT_{s}}\nonumber  \\
& + \sqrt{P_{s}}\sum_{p = 1}^{\widetilde{P}}{{\widetilde{h}}_{p}s_{c}(n- {\widetilde{\epsilon}}_{p}})e^{j2\pi{\widetilde{f}}_{d,p}nT_{s}} \nonumber \\
& - \sum_{p = 1}^{\widetilde{P}}{\widehat{h}}_{p}\sqrt{{1 - P}_{s}}s_{r}\left( n- {\widehat{l}}_{p} \right)e^{j2\pi{\widehat{f}}_{d,p}nT_{s}}\  + \ \eta\left( n\right) .
\end{align}
At the ideal case of $\widehat{h}_p=\widetilde{h}_p$, $\widehat{l}_p=\widetilde{\epsilon}_p$, $\widehat{f}_{d,p}=\widetilde{f}_{d,p}$, (\ref{eq-SIC}) can be expressed as
\begin{equation}
y_{I}\left( n\right) = \sqrt{P_{s}}\sum_{p = 1}^{\widetilde{P}}{{\widetilde{h}}_{p}s_{c}(n- {\widetilde{\epsilon}}_{p}})e^{j2\pi{\widetilde{f}}_{d,p}nT_s}\  + \ \eta(t).
\end{equation}

\subsection{The computational complexity} \label{complexity-analysis}
In the following, we provide a complexity analysis of the sensing-aided channel estimation, the interference cancellation, the FCCR sensing, and the DMD algorithm.

In \eqref{eq-ref-signal}, regenerating the reference signal for the pth significant path needs $8MN_a$ real multiplication and $4MN_a$ real addition. In \eqref{eq-first-path}, calculating the channel coefficient needs $6MN_a+4$ real multiplication and $5MN_a$ real addition. In \eqref{eq-re-gen-path},  generating the signal corresponding to the path needs $12MN_a$ real multiplication and $6MN_a$ real addition. Canceling interference needs $2MN_a$ real addition. In summary, $26MN_a + 4$ real multiplications and $17MN_a$ real additions are needed to calculate the channel coefficient of one effective path. Since there are $\widetilde{P}$ effective paths, sensing-aided channel estimation needs $\widetilde{P}(26M{N}_{a}  + 4)$ real multiplication and $17\widetilde{P}M{N}_{a}$ real addition.  

The FCCR algorithm requires $4MN{\mathrm{log}}_{2}N+4MN+2N{M}_{1} \mathrm{log}_{2} {M}_{1}$ real multiplication and $6MN\mathrm{log}_{2} N+2MN+4N{M}_{1} \mathrm{log}_{2} {M}_{1}$ real addition.

In the DMD algorithm, after down-sampling, the number of samples per chirp becomes \(N_{D} = (N - 2N_{cp})/D\).  The DMD algorithm needs $2MN_{D}{\mathrm{log}}_{2}N_{D}+2MN_{D}+2N_{D}{M}_{1} \mathrm{log}_{2} {M}_{1} + 2N_{D}M_{1}+2MN_{a}N_{F}+4MN_{a}$ real multiplication and $4MN_{D}{\mathrm{log}}_{2}N_{D}+4N_{D}{M}_{1} \mathrm{log}_{2} {M}_{1} + 2MN_{a}N_{F}+MN_{a}$ real addition.

The interference cancellation needs $12\widetilde{P}M{N}_{a}$ real multiplication and $8\widetilde{P}MN_a$ real addition. The computational complexity is summarized in Table~\ref{analysis}.

 \begin{table}[h] 
	\captionsetup{justification=centering}
	\caption{The computational complexity}
	\centering
	\begin{tabular}{|p{1.7cm}|p{3.0cm}| p{3.0cm}|}
\hline
\textbf{Algorithm}  & \textbf{Real multiplication}  & \textbf{Real addition}\\
\hline
Sensing-aided channel estimation & $\widetilde{P}(26M{N}_{a}  + 4)$  &  $17\widetilde{P}M{N}_{a}$\\
\hline
FCCR sensing &  $4MN{\mathrm{log}}_{2}N+4MN+2N{M}_{1} \mathrm{log}_{2} {M}_{1}$ & $6MN\mathrm{log}_{2} N+2MN+4N{M}_{1} \mathrm{log}_{2} {M}_{1}$ \\
\hline
DMD sensing  & $2MN_{D}{\mathrm{log}}_{2}N_{D}+2N_{D}{M}_{1} \mathrm{log}_{2} {M}_{1} + 2N_{D}M_{1}+2MN_{a}N_{F}+2MN_{D}+4MN_{a}$ &  $4MN_{D}{\mathrm{log}}_{2}N_{D}+4N_{D}{M}_{1} \mathrm{log}_{2} {M}_{1} + 2MN_{a}N_{F}+MN_{a}$ \\
\hline
Interference cancellation &  $12\widetilde{P}M{N}_{a}$ & $8\widetilde{P}MN_a$\\
\hline
\end{tabular}
	\label{analysis} 
\end{table}

\section{Performance Evaluations} \label{performance-evaluations}
In this section, we evaluate the performance of the Co-FMCW-OFDM scheme. For simplicity, we only consider two targets, the reflection power of the two targets is 0 dB and -6 dB, respectively. The simulation parameters are shown in Table~\ref{tb-parameter}.

 \begin{table}[h] 
	\captionsetup{justification=centering}
	\caption{Simulation Parameters}
	\centering
	\begin{tabular}{|p{4.5cm}|p{3.6cm}|}
\hline
\textbf{Parameters}  & \textbf{Values} \\
\hline
Number of OFDM subcarriers & $N=4096$ \\
\hline
Number of subcarriers to carrier data  & \(N_{sc}=3112\) \\
\hline
Length of CP  & \(N_{cp}=288 \)\\
\hline
Subcarrier spacing & \(\mathrm{\Delta}f=15\) KHz \\
\hline
Channel bandwidth & 50 MHz \\
\hline
Sampling rate & 61.44 MHz \\
\hline
Carrier frequency  & \(f_{c}=23.6\) GHz \\
\hline
Modulation scheme & QPSK \\
\hline
LDPC code & (1944, 972), coding rate 0.5 \\
\hline
Power weight for OFDM signal & $P_s=0.8930$ \\
\hline
Number of OFDM/FMCW symbols & $M=140$ \\
\hline
Relative reflection power of two targets & [ 0 -6 ] dB \\
\hline
Range of first target & Uniformly distributed in [48.8 244.14] m \\
\hline
Range of second target & Uniformly distributed in [146.48 341.79]
m \\
\hline
\end{tabular}
	\label{tb-parameter} 
\end{table}

\subsection{Sensing Performance Evaluations}\label{sensing-performance-evaluations}
We compare the FCCR sensing and the DMD sensing under two scenarios: 1) scenario a, the speed of the first and second target is uniformly distributed in
[45.7 91.5] km/hr, and the corresponding Doppler
frequency is uniformly distributed in [1.0 2.0] KHz. Note that in
this scenario, the ratio of average Doppler frequency to subcarrier
spacing is 10\(\%\). 2) scenario b, the speed of first and second
target is uniformly distributed in [228.8 274.5] km/hr, the
corresponding Doppler frequency is uniformly distributed in [5.0
6.0] KHz. Note that, in this scenario, the ratio of average Doppler
frequency to subcarrier spacing is 36.67\(\%\). 
According to the setting, the theoretic range and speed estimation resolution are
$R_{r} = \frac{cT_s}{2} = 2.44\ m$ and
$R_{v} = \frac{c}{2MM_1T_{sym}f_{c}}$ = 0.068 \(m/s\), respectively \cite{Zeng2020-7}.

\begin{figure}[!]
	\centering
	\includegraphics[width=\linewidth]{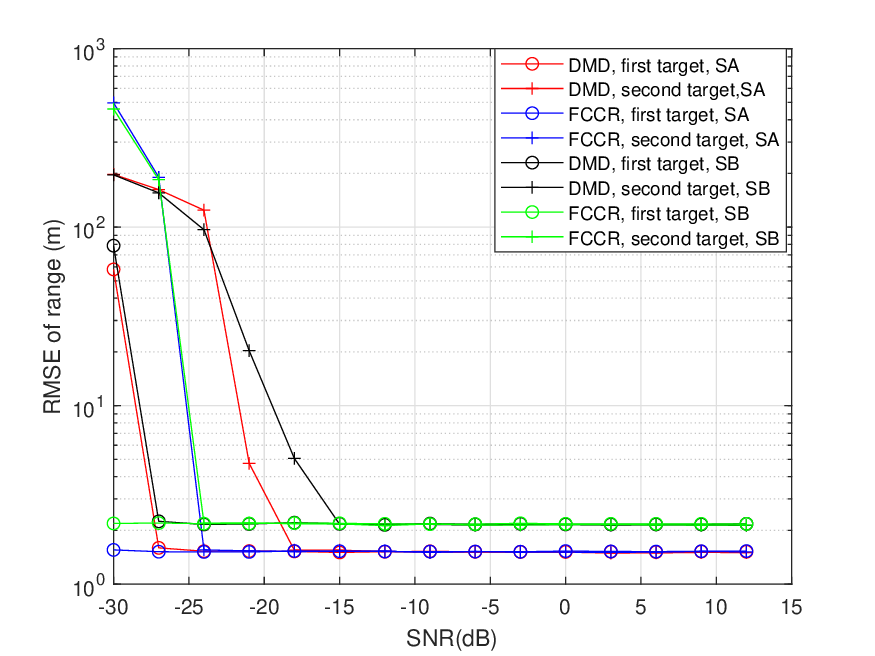}
\caption{Comparison of range RMSE between DMD sensing and FCCR sensing}
	\label{fig6}
\end{figure}

\begin{figure}[!]
	\centering
	\includegraphics[width=\linewidth]{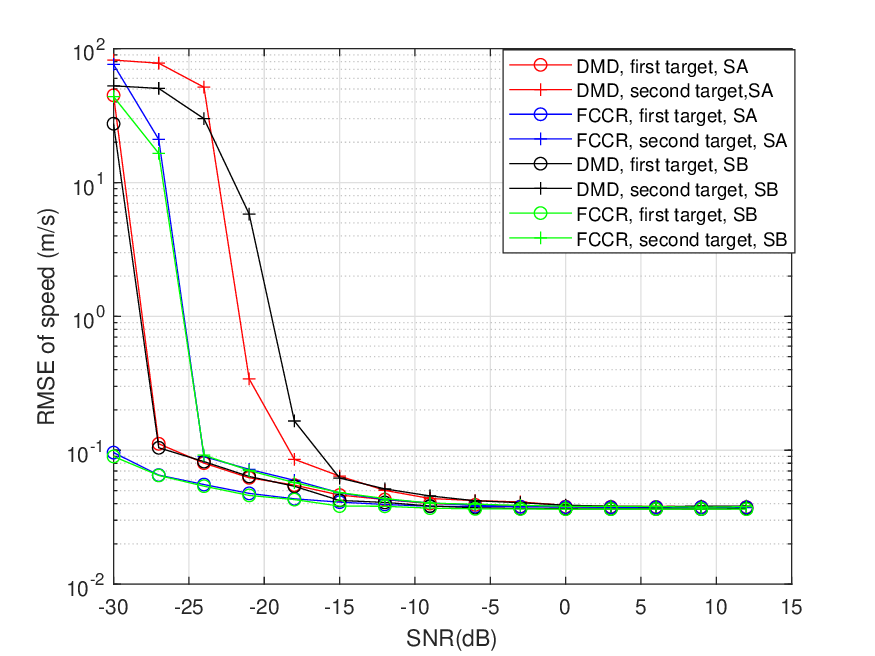}
\caption{Comparison of speed RMSE between DMD sensing and FCCR sensing}
	\label{fig7}
\end{figure}

Fig.~\ref{fig6} and Fig.~\ref{fig7} compare the RMSE of range and speed estimations between the FCCR and DMD sensing under scenario a (denoted by SA) and scenario b (denoted by SB). Under both scenarios,
FCCR sensing outperforms DMD sensing by 3 dB for the range and speed
estimation. We can see that under both scenarios, at high SNR, the RMSE
of speed is 0.036 \(m/s\), which is approximately half of the speed
resolutions. In scenario a, where the Doppler frequency is relatively
low, the RMSE of range at high SNR is 1.46 \(m\). In scenario b, where
the Doppler frequency is relatively high, the RMSE of range at high SNR
is 2.13 \(m\). Therefore, when Doppler frequency is low, FCCR and DMD
sensing achieves a range accuracy close to half of the range resolution
at high SNR. However, when Doppler frequency is high, FCCR and DMD
sensing achieves a range accuracy larger than half of the range
resolution. This is because, at high Doppler frequency, the approximations used in the FCCR sensing are not met exactly \cite{Zeng2020-7}, leading to degraded
sensing performance. 

\begin{figure}[!]
	\centering
	\includegraphics[width=\linewidth]{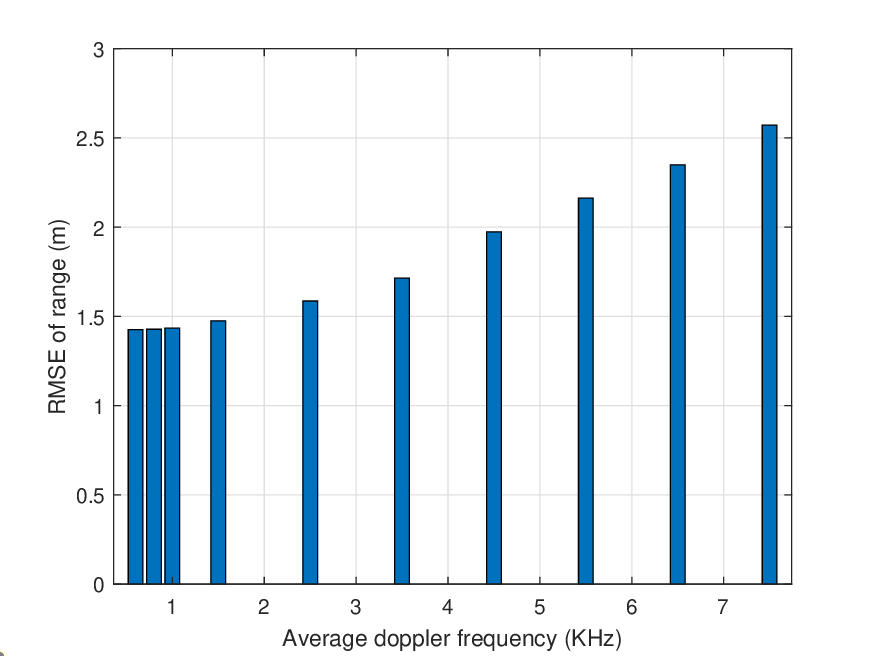}
\caption{Range RMSE versus Doppler frequency}
	\label{fig8}
\end{figure}

 To illustrate the impact of Doppler frequency on the RMSE of range, Fig.~\ref{fig8}  shows the RMSE of range at 12 dB SNR as a function of average Doppler   frequency. We can see that as the average Doppler frequency increases, the  RMSE of range also increases, indicating a sensing accuracy degradation. When the average Doppler frequency is increased from 1.5 KHz to 7.5KHz, the RMSE of range is degraded from 1.49 m to 2.6 m. 

 \subsection{Channel Estimation and Demodulation Performance Evaluations}\label{channel-estimation-performance-evaluation}
\begin{figure}[!]
	\centering
	\includegraphics[width=\linewidth]{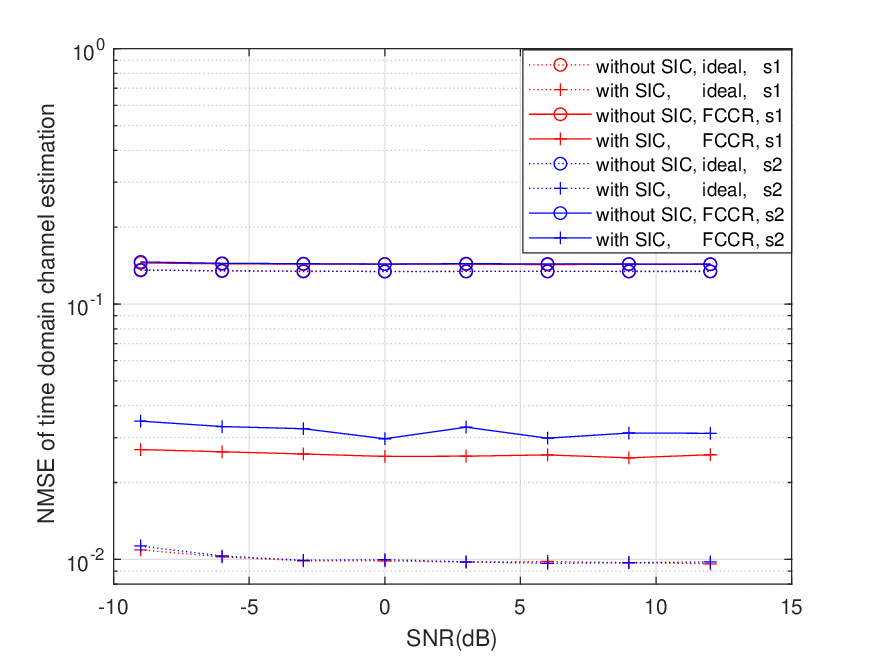}
\caption{NMSE of time domain channel estimation}
	\label{fig9}
\end{figure}

 In this subsection, we evaluate the proposed sensing-aided channel estimation performance and BER performance under two scenarios: 1) scenario 1 , the speed of the first and second target is uniformly distributed in [22.88  45.76] km/hr, and the corresponding Doppler frequency is uniformly distributed in [0.5 1.0] KHz. 2) scenario 2, the speed of the first and second target is uniformly distributed in [45.76  68.64] km/hr, the corresponding Doppler frequency is uniformly distributed in [1.0 1.5] KHz. 
 Fig.~\ref{fig9} shows the NMSE of the time domain channel estimation under scenario 1 (denoted by S1) and scenario 2 (denoted by S2). The NMSE of the time domain channel estimation is defined as
 \begin{equation}
\delta_{h,NMSE} = {\mathbb{E}\left\{ \frac{\sum_{p = 1}^{\widetilde{P}} \left| \widehat{h}_p - \widetilde{h}_p  \right|^{2}} {\sum_{p = 1}^{\widetilde{P}} \left| \widetilde{h}_p  \right|^{2}} \right\}}.
\end{equation}
In Fig.~\ref{fig9}, ``with SIC" denotes the performance of our proposed channel estimation with the SIC algorithm, as shown in subsection~\ref{sensing-aided-ce}. To demonstrate the advantage of SIC, we also show the channel estimation performance without SIC, denoted by ``without SIC”. ``FCCR” denotes the performance using FCCR sensing, while ``ideal” denotes the performance when sensing is ideal. Fig.~\ref{fig9} shows that the proposed channel estimation with SIC achieves better performance than channel estimation without interference cancellation. 

\begin{figure}[!]
	\centering
	\includegraphics[width=\linewidth]{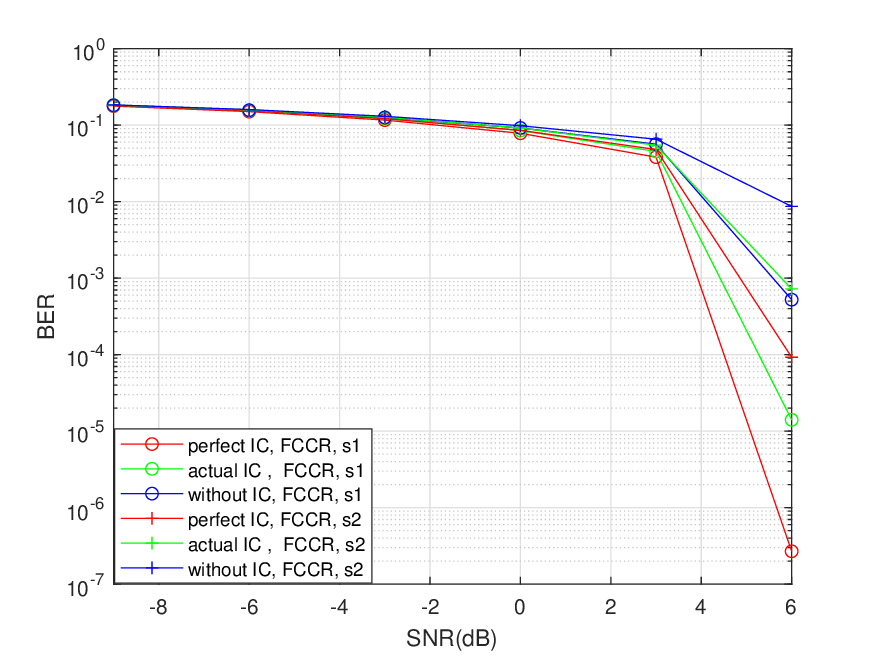}
\caption{Coded BER performances of the proposed system}
	\label{fig10}
\end{figure}

Fig.~\ref{fig10} illustrates the BER performance comparison among perfect FMCW interference cancellation (denoted by ``perfect IC”), actual FMCW interference cancellation (denoted by ``actual IC”) and without FMCW interference cancellation ( denoted by ``without IC”) under scenario 1 and scenario 2.  In this figure, FCCR is used for sensing. 
The gap between ``actual IC” and ``without IC” illustrates the gain obtained by FMCW interference cancellation. The gap between ``perfect IC” and ``actual IC” illustrates the performance loss due to non-perfect FMCW interference cancellation. The non-perfect FMCW interference cancellation is due to inaccuracies in the estimation of $\widehat{h}_p $, $\widehat{\tau }_p$ , and $ \widehat{f}_{d,p}$ causing the regenerated FMCW signal to differ from the actual FMCW signal. Consequently, the interference from the FMCW signal cannot be entirely eliminated.

As expected, BER performance degrades as the Doppler frequency increases. The reason for this is that the orthogonality among the OFDM subcariers are destroyed to some extent. Smaller Doppler frequency  leads to a larger gap between ``perfect IC" and ``actual IC". Smaller Doppler frequency also leads to a larger gap between ``actual IC" and ``without IC".

\begin{figure}[!]
	\centering
	\includegraphics[width=\linewidth]{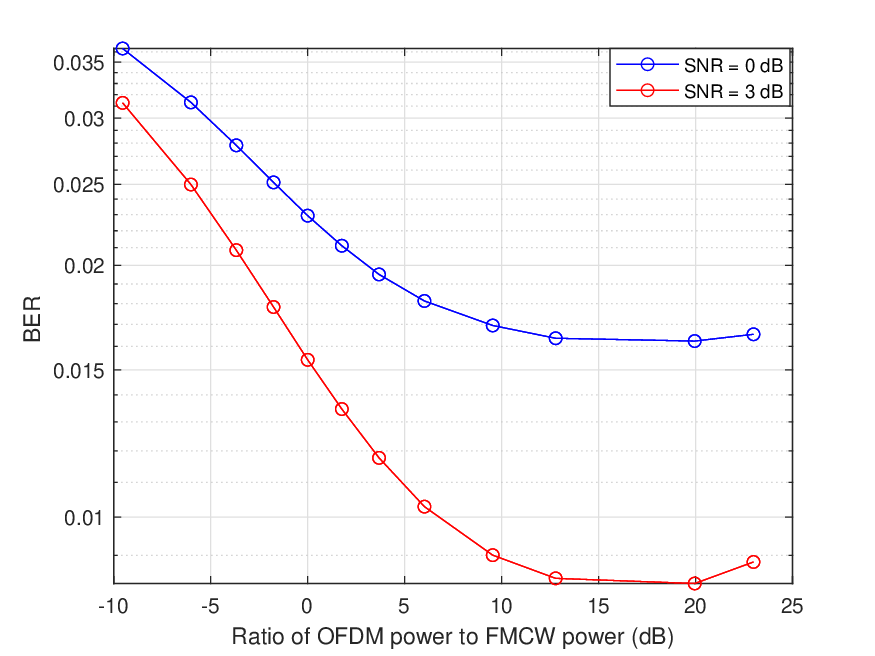}
\caption{Coded BER performances as a function of the power ratio between OFDM and FMCW signals}
	\label{figbervsps}
\end{figure}

 \subsection{ Impact of $P_s$ on communication BER performance}\label{BER vs $P_s$ }   
In the Co-FMCW-OFDM scheme, $P_s$, which represents the power allocated to OFDM signal, is a key design parameter. Increasing $P_s$ improves communication SNR but degrades the SNR for sensing and channel estimation. Fig.~\ref{figbervsps} illustrates the coded BER performance as a function of the power ratio between OFDM  and FMCW signals, defined as $R=10{\mathrm{log}}_{10} (P_s/(1-P_s))$.  In the simulation, two targets are considered with reflection powers of 0 dB and -6 dB, respectively. Their Doppler frequencies are uniformly distributed in the range [0.5 1.0] KHz and the number of symbols used for sensing is M = 28.
As shown in Fig.~\ref{figbervsps}, increasing the OFDM-to-FMCW power ratio  decreases the power available for FMCW signals, thus degrading sensing and channel estimation performance. However, the communication SNR improves with higher $P_s$, leading to an improvement in BER until a point where further increases in $P_s$ begin to negatively impact BER performance as well.  
 Depending on the application requirements, $P_s$ can be tuned to balance sensing and communication performance. For instances, $P_s$ can be selected such that the sensing SNR remains above a required threshold while minimizing the communication BER. 
 
 \section{Performance comparison between the proposed system and conventional system} \label{performance-comparison}
 In this section, we first compare the performance of the proposed scheme with the conventional OFDM scheme employing embedded pilot subcarrier. Then we compare the proposed scheme with OFDM-plus-FMCW system \cite{Wang2024-8}. 
 
 \subsection{Comparison with the conventional OFDM system} 
For the conventional OFDM scheme, we consider a similar frame and slot structure as that of the 5G standard: each slot contains 14 OFDM symbols, and each frame consists of 10 slots. Within each slot of 14 OFDM symbols, there are 3 OFDM symbols with Demodulation Reference Symbols (DMRS), located at the OFDM symbol 3, 8, and 12. In each of the three OFDM symbols, half of the subcarriers
are used as DMRS (pilot subcarriers), while the remaining subcarriers carry user
data. The DMRS pilot subcarriers in adjacent DMRS OFDM symbols
alternate between even and odd subcarriers. In the conventional OFDM scheme, these DMRS pilot symbols are used for the channel estimation.


We evaluate the two schemes by ensuring that the overhead of DMRS in the
conventional OFDM scheme is the same as the overhead of the FMCW signal
in the Co-FMCW-OFDM scheme. Since the conventional OFDM
scheme includes 3 DMRS OFDM symbols, with each symbol having only half of the
subcarriers as pilot subcarriers, the DMRS overhead is
calculated as 3/14/2 = 0.107. Consequently, the FMCW signal power is set
to \(1 - P_{s}\)= 0.107 in the Co-FMCW-OFDM scheme.

For the conventional OFDM scheme, we consider two methods for bistatic sensing: 1) channel estimation-based sensing and
2) decision feedback-based sensing. In the first method, the frequency domain channel is initially estimated with the DMRS pilots \cite{Sturm2009-3} as
\begin{equation}
\overline{H}\left( k_{1},m_{1} \right) = \frac{\overline{Y}\left( k_{1},m_{1} \right)}{b_{m_{1},k_{1}}},\ \left( k_{1},m_{1} \right) \in \ \Omega_{s}.
\end{equation}
Based on \(\overline{H}\left( k_{1},m_{1} \right)\)
(\(\left( k_{1},m_{1} \right) \in \ \Omega_{s}\)), a linear interpolation \cite{Baquero2019-4} is used to obtain the
channel estimation \(\widetilde{H}(k,m)\) (\(k = 0,1,\cdots,N - 1\),
\(m = 0,\ 1,\ \cdots,\ M - 1)\) for OFDM symbol \(m\) and
subcarrier \(k\),
\begin{equation}
\widetilde{H}(k,m) = \overline{H}\left( k,m_{1} \right) + \frac{\ \overline{H}\left( k,m_{2} \right) - \overline{H}\left( k,m_{1} \right)}{m_{2} - m_{1}} \left( m - m_{1} \right).
\end{equation}
Based on the estimated frequency domain channel, the well-known OFDM radar is used for the range and Doppler estimation  \cite{Sturm2009-3,Baquero2019-4,zeng2018-9}.

In the decision feedback-based sensing, the receiver first performs
demodulation and decoding based on the estimated channel described above to obtain the decoded payload data. The decoded payload data is then re-encoded,
modulated and mapped onto OFDM data subcarriers. This regenerated OFDM
signal is used as known signals or ``Pseudo-RS'' for FCCR sensing. For a
fair comparison, we also use FCCR sensing for the Co-FMCW-OFDM scheme.

\begin{figure}[!]
	\centering
	\includegraphics[width=\linewidth]{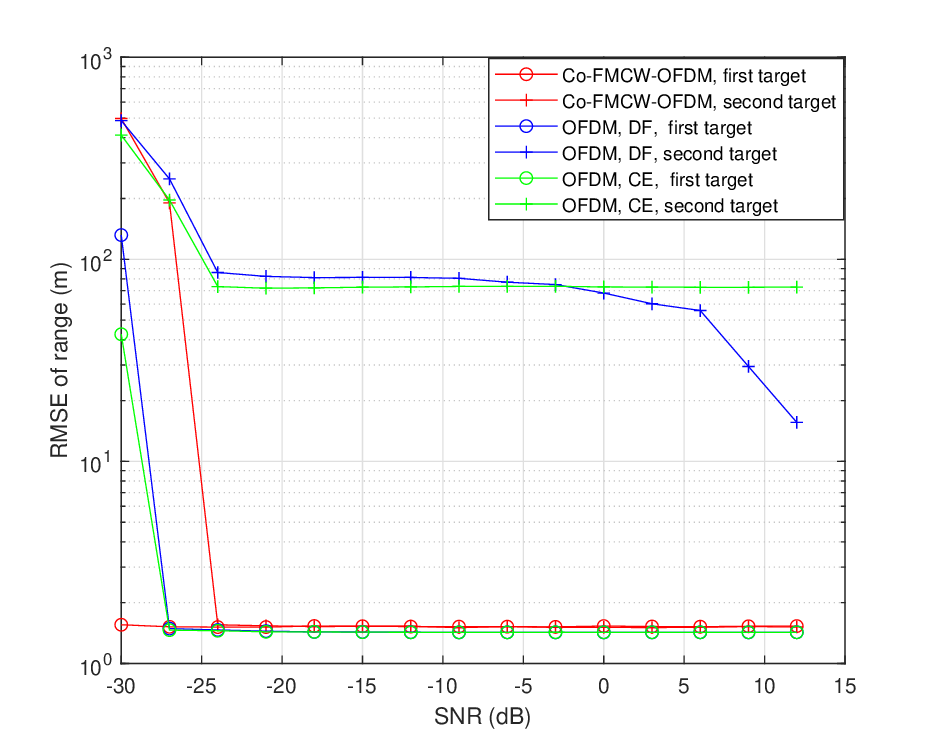 }
\caption{Comparison of range RMSE between Co-FMCW-OFDM and the conventional OFDM (scenario 2)}
	\label{fig12}
\end{figure}

\begin{figure}[!]
	\centering
	\includegraphics[width=\linewidth]{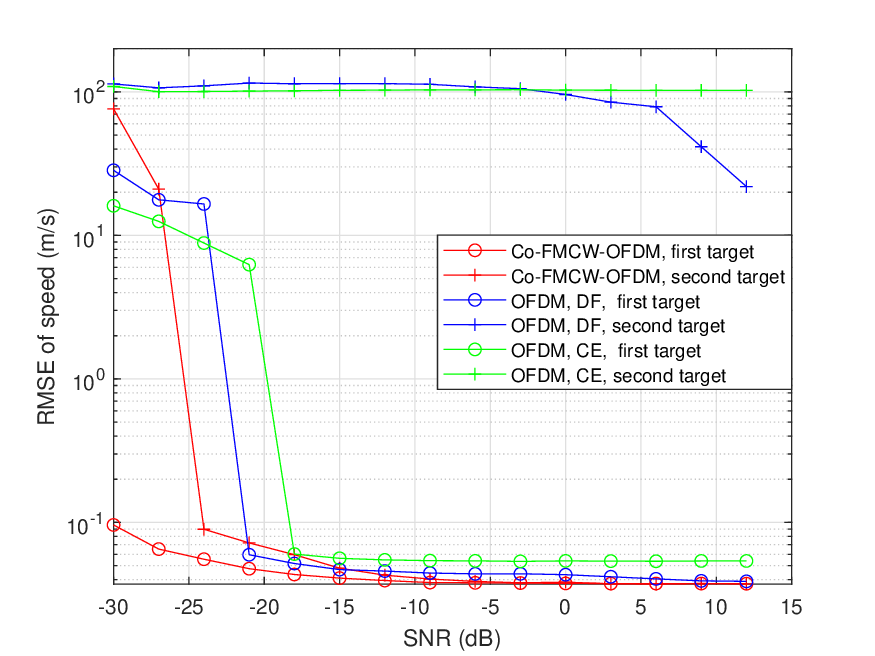}
\caption{Comparison of speed RMSE between Co-FMCW-OFDM and the conventional OFDM (scenario 2)}
	\label{fig13}
\end{figure}

First, we compare the sensing performances under scenario 1 and scenario 2 defined in Section~\ref{channel-estimation-performance-evaluation}. The simulation parameters are shown in Table~\ref{tb-parameter}. Fig.~\ref{fig12} and Fig.~\ref{fig13} compare the RMSE of the range and speed estimations, respectively, at scenario 2. Due to the page limit, we do not show the simulation results at scenario 1. In the figures, the performance of decision feedback-based sensing and channel estimation-based sensing are denoted by ``DF" and ``CE", respectively. We see that, for target 1, the Co-FMCW-OFDM scheme outperforms the OFDM scheme by 3 dB. For target 2, the Co-FMCW-OFDM scheme demonstrates a significant advantage over the OFDM scheme. This phenomenon can be explained as follows: since there are only 3 DMRS symbols within an OFDM frame, interpolation is used to estimate the channel across the entire grid. However, when the Doppler frequency is high, the channel changes rapidly. Thus, the interpolation cannot provide accurate channel estimations for the complete grid, leading to degraded sensing performance in channel estimation-based sensing schemes. 
In scenario 2, the Doppler frequency is high, and the orthogonality among OFDM subcarriers is destroyed to some extent. The combined effect of degraded channel estimation performance and destroyed orthogonality results in large errors in the decoded payload data. Using erroneous decoded payload data as ``pseudo-RS" will degrade the decision feedback-based sensing performance. 

\begin{figure}[!]
	\centering
	\includegraphics[width=\linewidth]{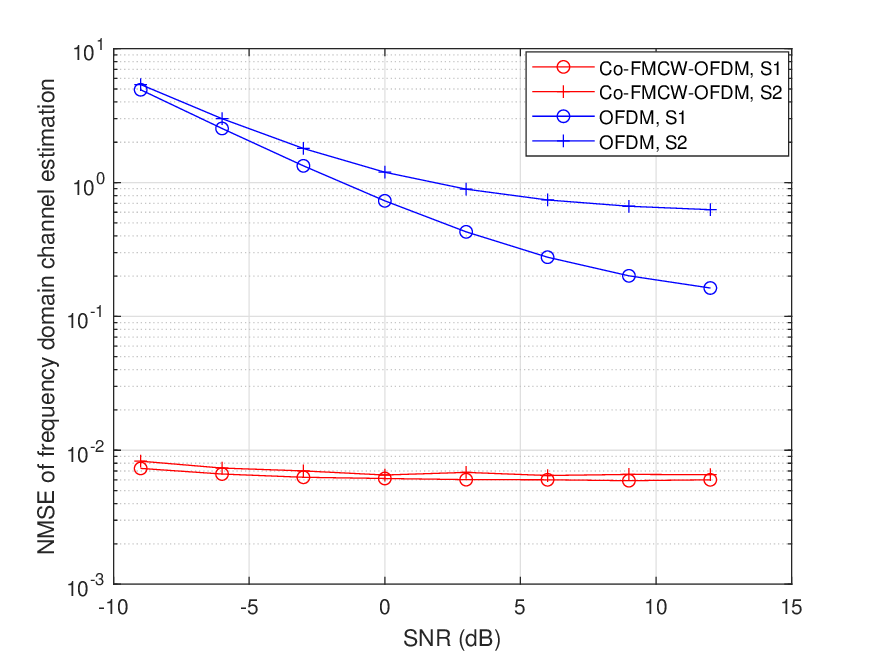}
\caption{ Comparison of channel estimation performance between Co-FMCW-OFDM and the conventional OFDM}
	\label{fig14}
\end{figure}

Fig.~\ref{fig14} compares the NMSE of the frequency domain channel estimations under scenario 1 (denoted by S1) and scenario 2 (denoted by S2). We see that the proposed scheme achieves much better performance than the conventional OFDM scheme. The reason is that, in the conventional OFDM scheme, the channel estimation for the entire grid is obtained through interpolation, which results in inaccurate channel estimation for fast time-variant channels. The higher the Doppler frequency, the worse the channel estimation accuracy. While in the proposed scheme, the superimposed pilot FMCW signal is effectively used to accurately estimate the delays, the Doppler frequencies, and the time domain channel coefficients, which are then utilized to generate more accurate frequency domain channel estimations for the entire grid. This is especially helpful for fast time-variant channels. We can also see that, as the Doppler frequency becomes larger, the performance gap between the proposed scheme and the conventional OFDM scheme becomes more obvious.

\begin{figure}[!]
	\centering
	\includegraphics[width=\linewidth]{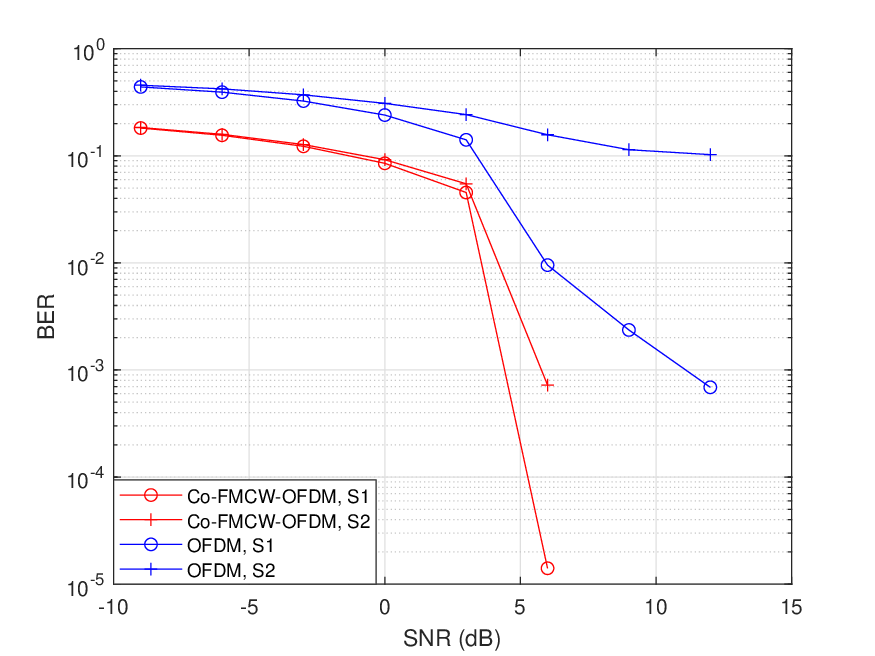}
\caption{ Comparison of BER between Co-FMCW-OFDM and the conventional OFDM}
	\label{fig15}
\end{figure}

Fig.~\ref{fig15} compares the BER performances. At both scenarios, the proposed scheme achieves significantly better BER performance than the conventional OFDM scheme. The reason is the same as described above. 

 \subsection{Comparison with the OFDM-plus-FMCW system} 
In the following we present a comparison of the BER performance between the Co-FMCW-OFDM and the OFDM-plus-FMCW in \cite{Wang2024-8}. The simulation scenario involves a single target with the SNR of 10 dB. The target’s Doppler frequency is uniformly distributed in the range [0.5, 1.0] kHz. We evaluate two cases where the number of symbols used for sensing is $M = 56$ and $M = 140$.
Since \cite{Wang2024-8} does not provide a channel estimation algorithm, we assume an ideal frequency-domain channel for both OFDM demodulation and FMCW interference cancellation in their scheme. In contrast, our method employs the proposed sensing-aided channel estimation for FMCW interference regeneration and cancellation.

\begin{figure}[!]
	\centering
	\includegraphics[width=\linewidth]{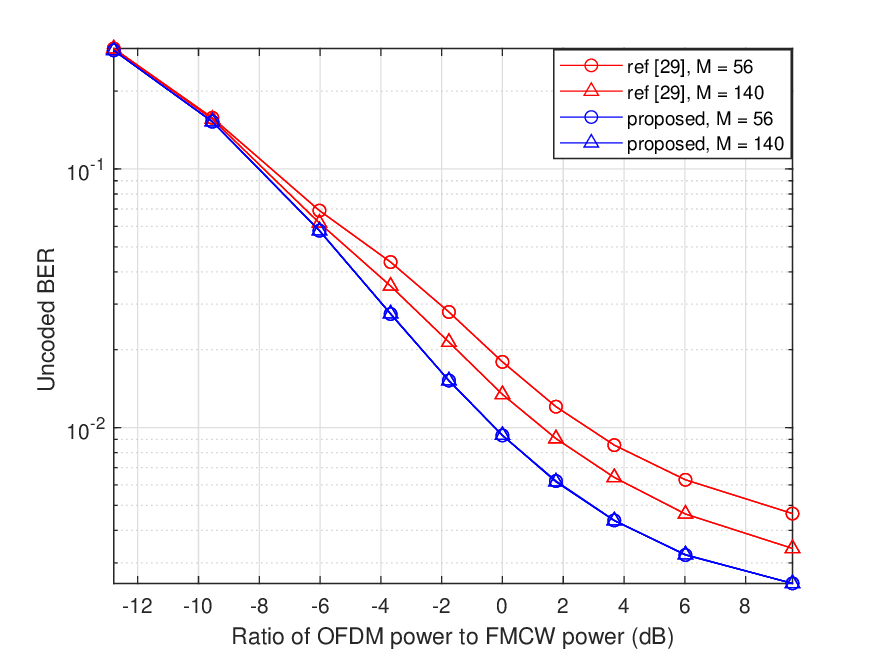}
\caption{Comparison of BER between Co-FMCW-OFDM and OFDM-plus-FMCW}
	\label{com_ref_27}
\end{figure}

As shown in Fig.~\ref{com_ref_27}, the simulation results show that the Co-FMCW-OFDM consistently achieves lower uncoded BER, even when we use an {\bf estimated channel}, whereas \cite{Wang2024-8} adopts an {\bf ideal channel}, for interference cancellation. As expected, increasing $M$ improves the performance of \cite{Wang2024-8} due to enhanced accuracy in the interference estimation via $M$ symbol averaging. However, in our scheme, the FMCW interference is regenerated and cancelled directly in the time domain, making it less sensitive to change of $M$.

\section{CONCLUSION} \label{conclusion}
We have proposed the Co-FMCW-OFDM for ISAC. In Co-FMCW-OFDM, the FMCW signal serves dual purposes: facilitating sensing and enabling channel estimation at the receiver end. We have proposed the FCCR and DMD bistatic sensing algorithms to estimate the delays and Doppler frequencies of all targets based on the FMCW pilots. With the estimated delays and Doppler frequencies, we have proposed a sensing-aided channel estimation based on the practical channel model and SIC, which achieves super estimation performance. Furthermore, a method has been presented to cancel the interference from the FMCW signals at the receiver prior to the OFDM demodulation. We have conducted comparative analyses and simulations on the sensing, channel estimation, and data demodulation performances. Simulation results show that the proposed system exhibits superior sensing, channel estimation and BER performances compared
to the conventional OFDM system and the OFDM-plus-FMCW in \cite{Wang2024-8}. 

%

\bibliographystyle{IEEEtran}
\bibliography{co-FMCW-OFDM}    

\begin{thebibliography}{10}
\providecommand{\url}[1]{#1}
\csname url@samestyle\endcsname
\providecommand{\newblock}{\relax}
\providecommand{\bibinfo}[2]{#2}
\providecommand{\BIBentrySTDinterwordspacing}{\spaceskip=0pt\relax}
\providecommand{\BIBentryALTinterwordstretchfactor}{4}
\providecommand{\BIBentryALTinterwordspacing}{\spaceskip=\fontdimen2\font plus
\BIBentryALTinterwordstretchfactor\fontdimen3\font minus
  \fontdimen4\font\relax}
\providecommand{\BIBforeignlanguage}[2]{{%
\expandafter\ifx\csname l@#1\endcsname\relax
\typeout{** WARNING: IEEEtran.bst: No hyphenation pattern has been}%
\typeout{** loaded for the language `#1'. Using the pattern for}%
\typeout{** the default language instead.}%
\else
\language=\csname l@#1\endcsname
\fi
#2}}
\providecommand{\BIBdecl}{\relax}
\BIBdecl

\bibitem{LiuFan2020-1}
F.~Liu, C.~Masouros, A.~P. Petropulu, H.~Griffiths, and L.~Hanzo, ``Joint radar
  and communication design: Applications, state-of-the-art, and the road
  ahead,'' \emph{IEEE Transactions on Communications}, vol.~68, no.~6, pp.
  3834--3862, 2022.

\bibitem{LiuFan2022-2}
F.~Liu, Y.~Cui, C.~Masouros, J.~Xu, T.~X. Han, Y.~C. Eldar, and S.~Buzzi,
  ``Integrated sensing and communications: Toward dual-functional wireless
  networks for {6G} and beyond,'' \emph{IEEE Journal on Selected Areas in
  Communications}, vol.~40, no.~6, pp. 1728--1767, 2022.

\bibitem{Hexa-X2023-13}
Hexa-X, ``Deliverable d3.3: Final models and measurements for localization and
  sensing,'' \url{https://hexa-x.eu/wp-content/uploads/2023/05}, 2023.

\bibitem{XTian2017-OFDM1}
X.~Tian, T.~Zhang, Q.~Zhang, and Z.~Song, ``Waveform design and processing in
  ofdm based radar-communication integrated systems,'' in \emph{Proc.
  {IEEE/CIC} Int. Conf. Commun. China ({ICCC})}, 2017, pp. 1--6.

\bibitem{MJNowak2016-fmcw1}
M.~J. Nowak, Z.~Zhang, Y.~Qu, D.~A. Dessources, M.~Wicks, and Z.~Wu,
  ``Co-designed radar-communication using linear frequency modulation
  waveform,'' in \emph{IEEE Mil. Commun. Conf.}, 2016, pp. 918--923.

\bibitem{Keskin2021-19}
M.~F. Keskin, V.~Koivunen, and H.~Wymeersch, ``Limited feedforward waveform
  design for {OFDM} dual-functional radar-communications,'' \emph{IEEE
  Transactions on Signal Processing}, vol.~69, pp. 2955--2970, 2021.

\bibitem{LiuY2017-20}
Y.~Liu, G.~Liao, J.~Xu, Z.~Yang, and Y.~Zhang, ``Adaptive {OFDM} integrated
  radar and communications waveform design based on information theory,''
  \emph{IEEE Communications Letters}, vol.~21, no.~10, pp. 2174--2177, 2017.

\bibitem{Keskin2020-21}
M.~F. Keskin, R.~F. Tigrek, C.~Aydogdu, F.~Lampel, H.~Wymeersch, A.~Alvarado,
  and F.~M. Willems, ``Peak sidelobe level based waveform optimization for
  {OFDM} joint radar-communication,'' in \emph{2020 17th European Radar
  Conference ({EuRAD })}, 2020, pp. 1--6.

\bibitem{LiChen2021-22}
L.~Chen, F.~Liu, W.~Wang, and C.~Masouros, ``Joint radar-communication
  transmission: A generalized pareto optimization framework,'' \emph{IEEE
  Transactions on Signal Processing}, vol.~69, pp. 2752--2765, 2021.

\bibitem{Haocheng2023}
H.~Hua, J.~Xu, and T.~X. Han, ``Optimal transmit beamforming for integrated
  sensing and communication,'' \emph{IEEE Transactions on Vehicular
  Technology}, vol.~72, no.~8, pp. 10\,588--10\,603, 2023.

\bibitem{BinLiao2023}
B.~Liao, X.~Xiong, and Z.~Quan, ``Robust beamforming design for dual-function
  radar-communication system,'' \emph{IEEE Transactions on Vehicular
  Technology}, vol.~72, no.~6, pp. 7508--7516, 2023.

\bibitem{XiangLiu2020}
X.~Liu, T.~Huang, N.~Shlezinger, Y.~Liu, J.~Zhou, and Y.~C. Eldar, ``Joint
  transmit beamforming for multiuser {MIMO} communications and {MIMO} radar,''
  \emph{IEEE Transactions on Signal Processing}, vol.~68, pp. 3929 -- 3944,
  2020.

\bibitem{XinHe2020}
X.~He and L.~Huang, ``Joint {MIMO} communication and {MIMO} radar under
  different practical waveform constraints,'' \emph{IEEE Transactions on
  Vehicular Technology}, vol.~69, no.~12, pp. 16\,342--16\,347, 2020.

\bibitem{Yingdu2023}
Y.~Du, Y.~Liu, K.~Han, J.~Jiang, W.~Wang, and L.~Chen, ``Multi-user and
  multi-target dual-function radar-communication waveform design: Multi-fold
  performance tradeoffs,'' \emph{IEEE Transactions on Green Communications and
  Networking}, vol.~7, no.~1, pp. 483--496, 2023.

\bibitem{LiuFan2021-17}
F.~Liu, Y.-F. Liu, A.~Li, C.~Masouros, and Y.~C. Eldar, ``Cramér-rao bound
  optimization for joint radar-communication beamforming,'' \emph{IEEE
  Transactions On Signal Processing}, vol.~70, pp. 240--253, 2021.

\bibitem{LiuFan2018-24}
F.~Liu, L.~Zhou, C.~Masouros, A.~Li, W.~Luo, and A.~Petropulu, ``Toward
  dual-functional radar-communication systems: Optimal waveform design,''
  \emph{IEEE Transactions on Signal Processing}, vol.~66, no.~16, pp.
  4264--4279, 2018.

\bibitem{otfs-2017}
R.~Hadani, S.~Rakib, M.~Tsatsanis, A.~Monk, A.~J. Goldsmith, A.~F. Molisch, and
  R.~Calderbank, ``Orthogonal time frequency space modulation,'' in \emph{2017
  IEEE Wireless Communications and Networking Conference (WCNC)}, 2017, pp.
  1--6.

\bibitem{PRav2019}
P.~Raviteja, K.~T. Phan, and Y.~Hong, ``Embedded pilot-aided channel estimation
  for {OTFS} in delay–doppler channels,'' \emph{IEEE transactions on
  vehicular technology}, vol.~68, no.~5, pp. 4906 -- 4917, 2019.

\bibitem{Linhai-2022}
H.~Lin and J.~Yuan, ``Orthogonal delay-{Doppler} division multiplexing
  modulation,'' \emph{IEEE Transactions on Wireless Communications}, vol.~21,
  no.~12, pp. 11\,024--11\,037, 2022.

\bibitem{Farhang-2023-ICC}
P.~S. Sanoopkumar and A.~Farhang, ``A practical pilot for channel estimation of
  {OTFS},'' in \emph{IEEE International Conference on Communications (ICC)},
  2023, pp. 1--6.

\bibitem{Tusha-2023}
A.~Tusha and H.~Arslan, ``Low complex inter-{Doppler} interference mitigation
  for {OTFS} systems via global receiver windowing,'' \emph{IEEE Transactions
  on Vehicular Technology}, vol.~11, no.~12, pp. 2670--2674, 2022.

\bibitem{Zhang-2023-ICC}
K.~Zhang, W.~Yuan, S.~Li, F.~Liu, F.~Gao, P.~Fan, and Y.~Cai, ``Radar sensing
  via {OTFS} signaling: A delay {Doppler} signal processing perspective,'' in
  \emph{IEEE International Conference on Communications (ICC)}, 2023, p.
  6429–6434.

\bibitem{Dazhuo2024}
D.~Wang, Y.~Zeng, Y.~Wang, F.~Chin, Y.~Ma, and S.~Sun, ``Pilot design, channel
  estimation, and target detection for integrated sensing and communication
  with {OTFS},'' in \emph{20th IEEE Asia Pacific Wireless Communications
  Symposium (APWCS 2024)}, 2024, pp. 1--6.

\bibitem{Sturm2009-3}
C.~Sturm, T.~Zwick, and W.~Wiesbeck, ``An {OFDM} system concept for joint radar
  and communications operations,'' in \emph{2009 IEEE 69th Vehicular Technology
  Conference ({VTC Spring 2009})}, 2009, pp. 1--6.

\bibitem{Baquero2019-4}
C.~B. Barneto, T.~Riihonen, M.~Turunen, L.~Anttila, M.~Fleischer, K.~Stadius,
  J.~Ryynänen, and M.~Valkama, ``Full-duplex {OFDM} radar with {LTE} and {5G
  NR} waveforms: Challenges, solutions, and measurements,'' \emph{IEEE
  Transactions on Microwave Theory and Techniques}, vol.~67, no.~10, pp.
  4042--4054, 2019.

\bibitem{Nata2023-5}
N.~K. Nataraja, S.~Sharma, K.~Ali, F.~Bai, and A.~F. Molisch, ``Bistatic
  vehicular radar with {5G-NR} signals,'' in \emph{2023 IEEE Global
  Communications Conference ({GLOBECOM 2023})}, 2023, pp. 1--6.

\bibitem{Zhang2023-6}
X.~Zhang, Y.~Ma, Y.~Zeng, S.~Sun, and Y.~Wang, ``Bistatic joint radar and
  communication with {5G} signal for range speed angle detections,'' in
  \emph{2023 IEEE 98th Vehicular Technology Conference ({VTC2023-Fall})}, 2023,
  pp. 1--6.

\bibitem{Zeng2020-7}
Y.~Zeng, Y.~Ma, and S.~Sun, ``Joint radar-communication with cyclic prefixed
  single carrier waveforms,'' \emph{IEEE Transactions on Vehicular Technology},
  vol.~69, no.~4, pp. 4069--4079, 2020.

\bibitem{Wang2024-8}
J.~Wang, Z.~Bai, J.~Lian, Y.~Guo, G.~Zhu, and Y.~Wang, ``A power-domain
  non-orthogonal integrated sensing and communication waveform design using
  {OFDM},'' \emph{IEEE Wireless Communications Letters}, vol.~13, no.~4, pp.
  984--988, 2024.

\bibitem{Arslan2020}
M.~M. Sahin and H.~Arslan, ``Multi-functional coexistence of radar sensing and
  communication waveforms,'' in \emph{IEEE 92nd Vehicular Tech. Conf.
  ({VTC2020-Fall})}, 2020, pp. 1--5.

\bibitem{ArslanOpenJournal2020}
------, ``Application-based coexistence of different waveforms on
  non-orthogonal multiple access,'' \emph{IEEE Open Journal of the
  Communications Society}, vol.~2, pp. 66--79, 2020.

\bibitem{Arslan2022}
E.~Memisoglu, M.~M. Sahin, and H.~Arslan, ``Multi-functional coexistence of
  radar sensing and communication waveforms,'' in \emph{IEEE Wireless Commun.
  Network. Conf. ({WCNC})}, 2022, pp. 1--6.

\bibitem{Aydogdu2019TransIT}
C.~Aydogdu, M.~F. Keskin, N.~Garcia, H.~Wymeersch, and D.~W. Bliss, ``Radchat:
  Spectrum sharing for automotive radar interference mitigation,'' \emph{IEEE
  Transactions on Intelligent Transportation Systems}, vol.~22, no.~1, pp.
  416--429, 2019.

\bibitem{Aydogdu2020SigProcMgz}
C.~Aydogdu, M.~F. Keskin, G.~K. Carvajal, O.~Eriksson, H.~Hellsten, and
  H.~Herbertsson, ``Radar interference mitigation for automated driving:
  Exploring proactive strategies,'' \emph{IEEE Signal Processing Magazine},
  vol.~37, no.~4, pp. 72--84, 2020.

\bibitem{Richards2022-10}
M.~A. Richards, \emph{Fundamentals of Radar Signal Processing}.\hskip 1em plus
  0.5em minus 0.4em\relax New York, NY, USA: McGraw Hill, 2022.

\bibitem{Venon2022-11}
A.~Venon, Y.~Dupuis, P.~Vasseur, and P.~Merriaux, ``Millimeter wave {FMCW
  RADARs} for perception, recognition and localization in automotive
  applications: A survey,'' \emph{IEEE Transactions on Intelligent Vehicles},
  vol.~7, no.~3, pp. 533--555, 2022.

\bibitem{zeng2018-9}
Y.~Zeng, Y.~Ma, and S.~Sun, ``Joint radar-communication: Low complexity
  algorithm and self-interference cancellation,'' in \emph{2018 IEEE Global
  Communications Conference ({GLOBECOM})}, 2018, pp. 1--6.

\bibitem{Gaudio2019-12}
L.~Gaudio, M.~Kobayashi, B.~Bissinger, and G.~Caire, ``Performance analysis of
  joint radar and communication using {OFDM and OTFS},'' in \emph{2019 IEEE
  International Conference on Communications Workshops ({ICC Workshops})},
  2019, pp. 73--76.

\end{thebibliography}

\end{document}